# Evolution Language Framework for Persistent Objects

Tetsuo Kamina[a] , Tomoyuki Aotani[b] , and Hidehiko Masuhara[c]

a   Oita University, Japan
b   Sanyo-Onoda City University, Japan
c   Institute of Science Tokyo (formerly Tokyo Institute of Technology), Japan

**Abstract**

**Context**  Multi-schema-version data management (MSVDM) is the database technology that simultaneously supports multiple schema versions of one database. With the technology, multiple versions of one software system can co-exist and exchange data even when the system's data structure evolves along with versions.

**Inquiry**  While there have been developed MSVDM theories and implementations for relational databases, they are not directly applicable to persistent objects. Since persistent objects are commonly implemented by means of object-relational mapping (OR-mapping), we need a right level of abstraction in order to describe evolution of data structures and translate data accesses in between different versions.

**Approach**  We propose a new evolution language consisting of a set of evolution operations, each denoting a modification of the source code and implicitly defining the corresponding modification to the database schema. Given the existence of multiple mapping mechanisms from persistent objects to databases, we designed the evolution language at two levels. At the abstract level, it handles scenarios such as refactoring and adding classes and fields. At the concrete level, we provide definitions for different mapping mechanisms separately, leveraging the existing database evolution language that supports MSVDM.

**Knowledge**  Our evolution language is designed to support existing evolution operations proposed in prior work. Additionally, it introduces support for operations related to class hierarchy changes, which are not covered by previous approaches. Using our proposal, two concrete mapping mechanisms, namely, a JPA-like mapping and signal classes, can be provided separately. Furthermore, our evolution language preserves program behavior and covers common evolution operations in practice.

**Grounding**  This work is supported by the formal definition of both the target abstract core language and the proposed evolution language, the formulation of several theorems demonstrating the soundness of our proposals, and the proofs of these theorems. Additionally, an empirical study was conducted to investigate the evolution histories of three open-source projects.

**Importance**  To the best of our knowledge, our proposal is the first evolution language for persistent objects that supports MSVDM. Moreover, it is the first evolution language defined at an abstract level. By defining mappings separately, we can apply it to a wide range of persistent object mechanisms built on top of SQL.

## The Art, Science, and Engineering of Programming

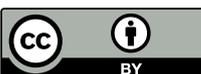
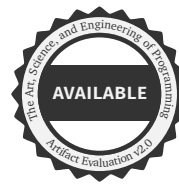
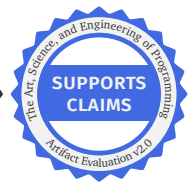

## 1 Introduction

Structures of objects and data are subject to change. When requirements for a software system change, we usually add/remove fields and data items to/from the existing object and data structures, and modify the relationship between them by introducing new objects, splitting and merging existing ones, and so on. We refer to this process as software evolution.

Evolution of a persistent object (an object whose state is kept in a database by some mapping mechanism from objects to the database) is challenging for two reasons: (1) persistent objects and their corresponding constructs in the database need to co-evolve. In some cases, this co-evolution should be performed implicitly, because the structures of database constructs are invisible to the persistent object. For example, signal classes [17] allow us to transparently connect to the corresponding database constructs, making the programmer unaware of the underlying databases; (2) in many cases, such evolution should consider scenarios in which the old version of the software system (based on the old object structures) is still actively used. In such a case, the old version should be able to access the database as if the database schema has not been changed, and the status of both versions of the database should be ensured to be "equivalent."

One notable existing work to address the former challenge is ESCHER [22], which supports a set of schema modification operations (SMOs) that propagate changes in persistent object structures to the database ones. However, this technique does not support the coexistence of multiple versions in a consistent manner. The latter challenge is strongly related to the multi-schema-version data management (MSVDM) technique [11], which enables multiple schema versions in a database system to coexist (Figure 1a). In this technique, the database schema evolution is managed by a database evolution language (DEL) that automatically generates the *delta code* that performs the propagation of a database update on a specific schema version to others *bidirectionally*. Specifically, Herrmann et al. proposed a DEL called BiDel, which consists of a set of SMOs that perform such delta code generation [11]. This technique is, however, limited to software systems that access the database through SQL.

The purpose of this paper is to propose a new evolution language with formal semantics that can be applied to persistent objects and support MSVDM. Similar to BiDel, our evolution language consists of a set of SMOs that translate persistent objects into evolved ones. We note that designing SMOs for persistent objects itself is challenging because the mapping from objects to the database has several variations (which are not considered by ESCHER). For example, the Java Persistence API (JPA) defines a standard API for object-relation mapping for Java. However, recently, persistent objects have also been studied in another programming paradigm, namely, reactive programming (RP), in which a different mapping strategy is applied [17].

*Contributions*. To address this problem, we design our evolution language at two levels. At the abstract level, we first define a core language as the target of our evolution language (Section 3). This core language is *abstract* in that the concrete mapping mechanisms can be defined independently. Based on this core language, we then define an *abstract* evolution language (Section 4). Among several evolution

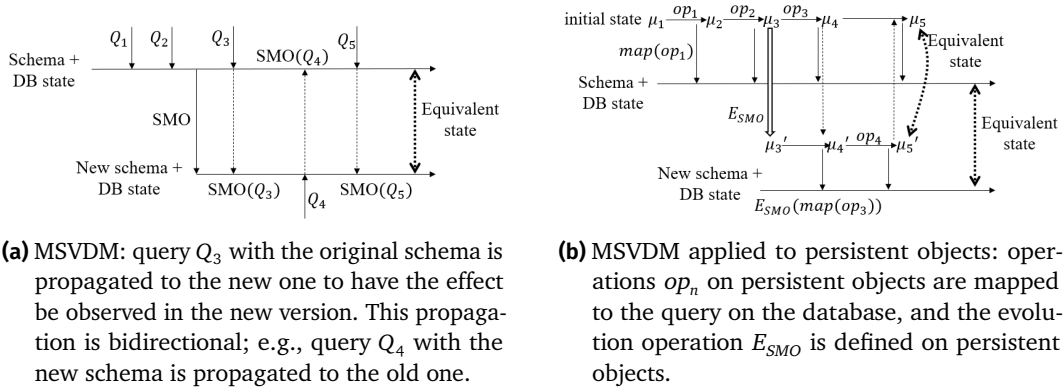

**(a)** MSVDM: query $Q_5$ with the original schema is propagated to the new one to have the effect be observed in the new version. This propagation is bidirectional; e.g., query $Q_4$ with the new schema is propagated to the old one.

**(b)** MSVDM applied to persistent objects: operations $op_n$ on persistent objects are mapped to the query on the database, and the evolution operation $E_{SMO}$ is defined on persistent objects.

■ **Figure 1** MSVDM and our proposal.

scenarios, this paper focuses on the scenarios related to the refactoring and addition of new elements. Specifically, our evolution language can describe the renaming of classes and fields, extraction of a superclass, and addition of classes and fields, as well as field type changes and field deletion. We note that this abstract evolution language is independent of MSVDM.

To operationalize this evolution language, a concrete definition is required. There are multiple ways to define such a concrete definition. Among these, we consider supporting MSVDM to be an important property that the evolution language should satisfy, as discussed above. Thus, at the concrete level, we provided the definitions of two mapping mechanisms, namely, JPA-like mapping and signal classes (the RP-style mapping mentioned above), leveraging BiDel's SMOs (Section 5). This means that our definitions support MSVDM, in which an update of persistent objects in a version is bidirectionally propagated to other versions (Figure 1b). For each of these mapping mechanisms, we also prove that the aforementioned assumptions hold.

To investigate the expressive power of the proposed evolution language, we conducted an empirical study targeting the evolution of real open-source JPA software (Section 6). By classifying the differences between declarations before and after a commit of each JPA class in the repository, we found that our evolution language covers a large part of existing evolution scenarios regarding persistent objects. We also identified some evolution scenarios that are not supported by our evolution language due to the expressiveness limitations of our language and BiDel. Although the study indicates that such evolution scenarios do not occur frequently, we consider that these limitations suggest a potential direction for future research.

All the definitions of the aforementioned proposals, namely, abstract core language, abstract evolution language, and two concrete mapping mechanisms, are formally provided. We also formulated and proved several theorems demonstrating the soundness of the proposals (Appendicies A, B, C, D, and E).

## 2 Background

This section introduces the prerequisite technologies for our proposal.

CREATE TABLE $C(\bar{f})$, $\mathscr{R} = \mathscr{R} \cup \{C \mapsto \pi_{\bar{f}}(\emptyset)\}$

RENAME TABLE $C$ INTO $D$, $\mathscr{R} = $ let x=$\mathscr{R}(C)$ in $(\mathscr{R} \cup \{D \mapsto x\}) \setminus \{C \mapsto x\}$

RENAME COLUMN $\bar{f}$ IN $C$ TO $\bar{g}$, $\mathscr{R} = \mathscr{R} \oplus \{C \mapsto \mathscr{R}(C)_{\bar{f}/\bar{g}}\}$

ADD COLUMN $\bar{f}$ AS $f(\bar{g})$ INTO $C$, $\mathscr{R} = \mathscr{R} \oplus \{C \mapsto \mathscr{R}(C)_{\bar{f} \leftarrow f(\bar{g})}\}$

DROP COLUMN $\bar{f}$ FROM $C$ DEFAULT $f(\bar{f})$, $\mathscr{R} = \mathscr{R} \oplus \{C \mapsto \mathscr{R}(C)_{\setminus \bar{f}}\}$

DECOMPOSE TABLE $C$ INTO $C_1(\bar{f}), C_2(\bar{g})$ FK $\bar{h}$, $\mathscr{R} = $
  let x=$\mathscr{R}(C)$ in $(\mathscr{R} \cup \{C_1 \mapsto \pi_{\bar{f}}(x)\} \cup \{C_2 \mapsto \pi_{\bar{g}}(x)\}) \setminus \{C \mapsto x\}$

JOIN TABLE $C_1, C_2$ INTO $C$, $\mathscr{R} = $
  $((\mathscr{R} \cup \{C \mapsto C_1 \bowtie C_2\}) \setminus \{C_1 \mapsto \mathscr{R}(C_1)\}) \setminus \{C_2 \mapsto \mathscr{R}(C_2)\}$

OUTER JOIN TABLE $C_1, C_2$ INTO $C$, $\mathscr{R} = $
  $((\mathscr{R} \cup \{C \mapsto C_1 \rhd\!\!\bowtie C_2\}) \setminus \{C_1 \mapsto \mathscr{R}(C_1)\}) \setminus \{C_2 \mapsto \mathscr{R}(C_2)\}$

■ **Figure 2** Definitions of SMOs in BiDel.

## 2.1 Multi-Schema-Version Data Management

The evolution of software systems occasionally requires changes in database schema; occasionally, this evolution requires multiple software versions to coexist at a time, resulting in the scenario in which multiple database schema versions also coexist. Keeping multiple database schema versions consistent is an error-prone and difficult task. This task forces developers to migrate a database with the older version to a new version completely. This migration requires a piece of manually written and maintained *delta code*, i.e., an implementation of propagation logic that is necessary to run an application with a schema version different to the version used by the database.

To solve this problem, a Multi-Schema-Version Data Management (MSVDM) system was proposed [11]. This system supports the creation, management, and deployment of different database schema versions, where all these versions coexist. A new schema is created from the existing one using a simple DEL, which consists of a set of SMOs. Each SMO automatically generates the delta code to ensure bidirectional data propagation of both forward and backward between schema versions.

**Database Evolution Language**  Herrmann et al. proposed the database evolution language BiDel [11], which consists of a set of SMOs. We summarize the syntax and semantics of the subset of SMOs on which our work is based in Figure 2. In this figure, we slightly modify the syntax to enforce each operation to take the set of relations $\mathscr{R}$ explicitly, as it is necessary to formally define the mapping mechanisms of persistent objects using BiDel (Section 5).

In these definitions, we adopt the following conventions. A relation in a relational database consists of its name and a set of tuples, and we represent this as a map from the name to the set of tuples. We apply the standard relational algebra notations [5];

e.g., $\pi_{\bar{f}}$ denotes a projection by the columns $\bar{f}$, and $\sigma_{cond}$ denotes a selection from a relation using the condition $cond$. $\mathscr{R}$ denotes the set of relations. Overlines denote sequences, e.g., $\bar{f}$ represents a possibly empty sequence $f_i, \cdots, f_n$, where $n$ is the length of the sequence. We assume that the name of a relation is unique in $\mathscr{R}$, and we often refer to $\mathscr{R}(\texttt{C})$ using only its name $\texttt{C}$. We write the tuples obtained by renaming $\bar{f}$ in $\mathscr{R}(\texttt{C})$ to $\bar{g}$ as $\mathscr{R}(\texttt{C})_{\bar{f}/\bar{g}}$, the tuples obtained by adding attributes $\bar{f}$ to $\mathscr{R}(\texttt{C})$ whose values are initialized by the function $f(\bar{g})$ as $\mathscr{R}(\texttt{C})_{\bar{f}\leftarrow f(\bar{g})}$, and the tuples obtained by dropping $\bar{f}$ from $\mathscr{R}(\texttt{C})$ as $\mathscr{R}(\texttt{C})_{\setminus \bar{f}}$. The binary operation $\mathscr{R}_1 \oplus \mathscr{R}_2$ denotes a destructive update, which results in a set of relations obtained by replacing all $\mathscr{R}_1(\texttt{C}_i)$ (where $\texttt{C}_i \in dom(\mathscr{R}_2)$) with $\mathscr{R}_2(\texttt{C}_i)$.

Each SMO is defined using an equation: the left-hand side defines the syntax, and the right-hand side defines a new schema after the SMO is issued; i.e., each equation defines the effect on the schema $\mathscr{R}$. This new schema is specified such that the relational schema is changed using set operations. Each SMO enables us to create or rename both tables and columns, following the generally known semantics of standard SQL-DDL. For example, `CREATE TABLE C(`$\bar{\texttt{f}}$`)`, $\mathscr{R}$ creates a relation $\texttt{C}$ with attributes $\bar{\texttt{f}}$ and adds that relation to $\mathscr{R}$. The `DECOMPOSE TABLE` operation vertically splits the table $\texttt{C}$ into two tables $\texttt{C}_1$ and $\texttt{C}_2$ by distributing the columns in $\texttt{C}$ into $\texttt{C}_1$ and $\texttt{C}_2$ using attributes $\bar{\texttt{h}}$ as foreign keys. Its inverse operation is provided as `JOIN TABLE`, which inner-joins two tables. If the keyword `OUTER` is provided, this operation performs the outer-join.

### 2.2 Evolution of Persistent Objects

The database schema for persistent objects (i.e., the mapping from objects to the database) is implicit and subject to change. Sometimes, such a mapping drastically changes according to the semantics of persistent objects even when their host languages are mostly identical. We illustrate that by introducing two different persistent objects mechanisms based on Java in the subsequent sections.

**Java Persistent API**   JPA is a standard API for object-relation mapping for Java. As this is an API for mapping records in a relational database system to Java objects and manipulating such Java objects, several implementations are available; among them, one of the simplest methods of mapping is to map a relational schema to a class, and each record stored in that relation to an instance of that class. In JPA, such a class is called an entity class, whose declaration is annotated by `@javax.persistence.Entity`. In an entity class, we can apply the annotation `@Id` for a field to specify that the field corresponds to the primary key (Figure 3).

For simplicity, we apply the following simplification throughout this paper. First, we consider that an entity class is implicitly bound with the relation with the same name. Second, without issuing an explicit operation to make the instance of an entity class persistent, we consider that an instance of an entity class has implicitly become persistent (if it already exists in the database, the instance is bound with that entry; otherwise, it inserts a new entry). We note that these simplifications make the underlying database system totally implicit. We assume that a constructor of an entity

```
1  @javax.persistence.Entity
2  class LoginStatus {
3       @Id String id;
4       boolean status;
5       LoginStatus(String id) { ... }
6  }
```

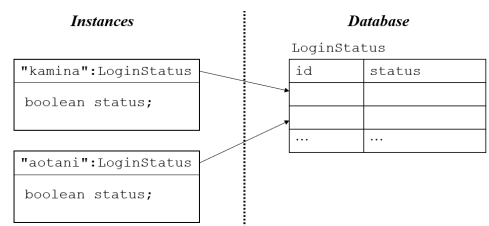

■ **Figure 3** An example of entity class (in our setting).

■ **Figure 4** Overview of JPA-like mapping

class always has the formal parameter id to specify a single primary key. An access to a field on an instance of an entity class results in a value of the column of the tuple identified by that primary key. We also assume that each entity class provides the set method that updates the fields (columns) values of the instance specified by the id. For example, the following code snippet updates the columns name and affiliation of the record identified by the key "kamina":

```
1  // getting the Author record with pKey "kamina"
2  LoginStatus kamina = new LoginStatus("kamina");
3  // updating the record values
4  a.setStatus(true);
```

Figure 4 depicts the object-relation mapping in this setting. The database table schema can be derived from the class declaration, and each instance corresponds to a tuple in that table. Each identifier is used as a primary key to identify that tuple.

**Signal Classes**    A signal class is a new language construct recently proposed by Kamina et al. [17] to make *signals* in reactive programming persistent. Signals are abstractions for time-varying values that can be declaratively connected to form dataflows; i.e., signals directly represent dataflows from inputs given by the environment to outputs that respond to the changes in the environment. Signals were first proposed in several functional languages such as Fran [10]; subsequently, several programming languages introduced their ability to imperatively change signals [6, 16, 20, 28]. Signal classes are introduced in SignalJ to group persistent signals (signals whose values are not transient but persistent such that all update histories are implicitly stored in the time-series database [15]) into a single module.

The following snippet shows an example of a signal class, which declares the signal class version of the login status class shown in Figure 3.

```
1  signal class LoginStatus {
2     persistent signal boolean status;
3     LoginStatus(String id) { ... } }
```

A signal class is declared using the modifier signal in the class declaration. This class declares a persistent signal status using the modifier persistent. This signal records the change history of the users login status as time-series data. We may also declare other signals that depend on this persistent signal; however, this feature is missing in

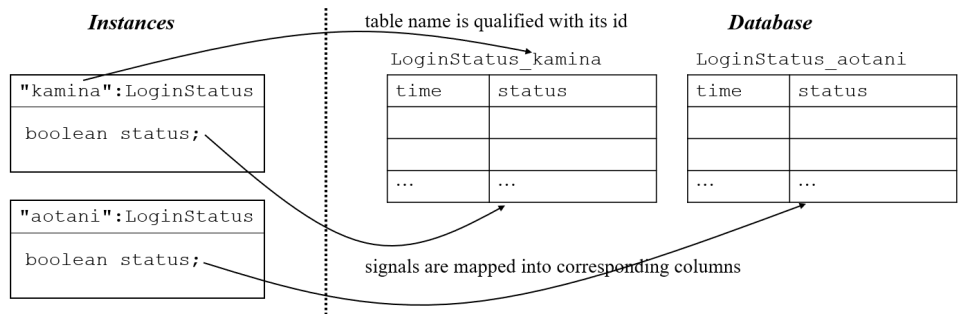

**■ Figure 5** Overview of signal class mapping this example. In SignalJ, update histories of signals that depend on persistent signals are also stored permanently.

Unlike JPA, each signal class instance records its update history as a time-series table, and a different mapping in which each instance corresponds to a table is applied. This table consists of a column time to store the timestamp when the tuple is inserted and other columns to store the values of corresponding persistent signals at the time specified by the timestamp. To distinguish two instances with different identifiers, each identifier is used as a part of the table name (Fig. 5).

**Research Objectives and Premises** We consider evolution of persistent objects. For example, we may add a field, namely, isPrivileged, to distinguish the privileged login users with other users in the next version. This evolution scenario applies to both persistent object mechanisms, JPA and signal classes. In general, regardless of the actual persistent object mechanisms, the evolution of *classes* should be identical.

This observation motivates us to design a new evolution language that can manage the evolution of such a different set of persistent object mechanisms uniformly. A fundamental challenge is how to abstract the differences between the persistent object mechanisms. For example, even though the evolution scenario is similar, signal classes significantly differ from JPA:

- The formal parameter id in the constructor does not represent a primary key but an identifier of the database table that records the update history of the specific signal class instance. For example, the constructor invocation new LoginStatus("kamina") implicitly bind the instance with the table identified by "kamina" (or generates a new table if the corresponding table does not exist).
- An access to a persistent signal results in a most recent value of the signal.
- Field updates in a signal class instance do not perform updating of record values but insert a new tuple with a new timestamp into the corresponding table.

Furthermore, these differences result in different mappings into SMOs in BiDel when considering MSVDM.

To address this issue, we define an evolution language framework for persistent objects where concrete definitions of persistent objects can be provided independently. This framework includes the following definitions:

- An abstract core language for persistent objects that accesses/modifies the database status using field accesses and constructor invocations, but how to access/modify the database is defined independently.
- An evolution language that manages the evolution of the abstract persistent objects using a uniform set of evolution operations. How each operation is mapped to existing SMOs is defined independently. With this mapping, this evolution language is applicable to any languages that conform to the abstract core language.

**Premises** Since our focus is on the co-evolution of persistent objects and database schema, this paper excludes modifications to programs unrelated to changes in the database schema, such as solely altering method definitions or adding transient fields. We design under the assumption that the proposed language is integrated into an editor, allowing users to choose an evolution scenario from a menu, after which the corresponding operation is executed internally. Unsupported modifications are directly made by editing the code using the editor.

This paper requires the host language providing persistent objects to be conformed with the aforementioned abstract core language that assumes relational database systems as the underlying databases. Other language and database mechanisms, such as JavaScript and JSON-based database systems, are beyond the scope of this paper.

In this paper, we examine software evolution scenarios that make different (new) versions of software systems. In MSVDM, various versions of a software system can coexist, with each version utilizing distinct versions of classes that share the same database. We do not address situations where different versions of the same class are running on the same Java process, and where running objects are modified online.

## 3 Abstract Core Language

In this section we introduce the syntax of abstract core language of persistent objects to which our evolution language is applied. To focus on the evolution language, its semantics and important theorems are presented in Appendices A and B, respectively. In the following, we assume a Java-like class-based language as the base of persistent objects. Furthermore, we assume that each mapping mechanism has the following properties:

- Each object (class) implicitly corresponds to a database table, and the class name is automatically bound with such a table.
- The states of the object (i.e., the values of the object's fields) are permanently stored in the underlying database system. For simplicity, we do not consider the transient fields whose values are not stored in the database.
- Each object is identifiable using the user-provided identifiers.

As we focus on evolution of class declarations, we define the core language of persistent object based on Featherweight Java (FJ) [14], which provides a *class table* to model class declarations. The syntax of the target core language is shown in Figure 6. Let the metavariables C, D, and E range over class names; f and g range over field

```
CL  ::=  class C extends D { C̄ f̄; M̄ }
M   ::=  C m(C̄ x̄) { return e; }
e   ::=  x | e.f | e.set(ē) | e.m(ē) | new C(l,ē) | l | l_c
```

■ **Figure 6** Abstract syntax of the target core language names; e ranges over expressions; x ranges over variables, which include a special variable this; l ranges over identifiers for database entities such as @ID on JPA classes; and m ranges over method names. We use $\bar{C}\ \bar{f}$ as shorthand for "$C_1\ f_1 \cdots C_n\ f_n$" and $\bar{C}\ \bar{f}{=}\bar{e}$ as shorthand for "$C_1\ f_1{=}e_1 \cdots C_n\ f_n{=}e_n$."

An expression can be either a variable, access to a field, invocation of a method (including a special method set), instance creation, identifier l, or identifier annotated with a class name $l_c$. We distinguish identifiers that appear in the source code and those in the runtime expressions, and the latter are always annotated with the class name.[1] We will see that this distinction is useful when considering the "extract superclass" refactoring, in which the identifier is duplicated into two relations. The set method is a predefined setter that takes arguments as many as the receiver's fields and sets them to the corresponding fields. Among these expressions, only l and $l_c$ are normal forms, i.e., every value is an identifier. These identifiers correspond to the *user-provided* identifiers for the persistent objects, and are required for constructor invocations. We assume the set *Id* of identifiers and $l \in Id$.[2]

For simplicity, we abstract away from the details of primitive values (and their corresponding types). A primitive value can simply be represented by a fresh location $l_U$ (U stands for *uninterpreted* base type) such that any field accesses and method invocation on that is prohibited. It is easy to extend the calculus with such primitive values, but for the discussions regarding persistent objects, only values with class types are relevant. Thus, this omission does not affect the subsequent technical discussions. Furthermore, our calculus does not model the method overriding, which we consider also does not affect the results presented in this paper.

A program (*CT*, e) in this core language consists of a class table *CT* that maps a class name C to a class declaration CL and an expression e that corresponds to the body of the main method. We also introduce typing relations, which are required to describe the semantics of our evolution language discussed in Section 4. A type environment $\Gamma$ is a finite mapping from variables to class names. A store environment $\Sigma$ is a finite mapping from identifiers to class names. A typing judgment for expressions is of the form $\Gamma \mid \Sigma \vdash \mathsf{e} : \mathsf{C}$, read as "expression $\mathsf{e}$ is given type $\mathsf{C}$ under the type environment $\Gamma$ and store environment $\Sigma$." We assume that $\Sigma$ is formerly constructed to be consistent with the program and the database $\mathscr{R}$. This construction is possible as $\mathscr{R}$ is given in our calculus, and we assume that all the identifiers are explicitly provided by the programmer (Section 2.2). For precise explanations, readers may refer to Appendix A.

---

[1] We assume that an identifier is "tagged" with the class name when it is used in the computation. This assumption is valid in signal classes in which the identifier is always tagged with the class name. This is also valid in JPA in which an access to $l_c$ can be implemented as SELECT * FROM C where id=l.

[2] This "a single key" assumption is made just for simplifying discussion. We do not consider that this simplification loses generality; e.g., multi-column keys can be represented by a simple extension of the syntax to support an instance creation C($\bar{l}$,$n$,$\bar{e}$), where $n$ is the number of $\bar{l}$.

## 4    Evolution Language Framework

Now, we consider the evolution of the programs written in the core language defined in Section 3. As mentioned earlier, in this paper we focus on evolution scenarios involving refactoring and addition of new elements. Lots of refactorings have been proposed to date, and many of them can be automated. Some refactorings, such as renaming and extracting a superclass, are relevant to schema evolution. For example, renaming fields and classes corresponds to renaming columns and tables, respectively, while extracting a superclass corresponds to splitting database tables. Other modifications, such as adding fields, are also relevant to schema evolution (e.g., corresponding to the addition of columns). Several refactorings (e.g., extracting a method) are however not relevant to the database schema evolution and thus not necessary to be discussed here. Referring to existing SMOs for persistent objects [22], we focus on the following operations:

- Creating a new class declaration.
- Renaming a class/fields in a class declaration.
- Adding/Deleting fields into/from a class declaration.
- Changing field types in a class declaration.
- Extracting a superclass.
- Merging a class declaration into its direct subclass.

In addition to SMOs in [22] (tailored to FJ), our SMOs include operations regarding class hierarchy changes (the last two in the above list), which are not supported by [22]. This is because, the representation of class hierarchies is important in our study, where different mapping strategies of inheritance can be considered. This will be further elaborated in Section 5. To automate these operations, it is necessary to formally define the semantics of them, i.e., the precise definition of the program after performing the operation, and the premise where we can safely issue that operation.

In this section, we develop an evolution language similar to BiDel but its target is now a Java-like language. Our objectives are (1) to support the schema evolution in which multiple versions coexist, allowing updates on one version to be consistently visible from other versions in a bidirectional manner, similar to BiDel, (2) to cause the evolution of the underlying database to be implicitly triggered when the evolution operation is applied to the source code, and (3) to support several persistent object mechanisms discussed in Section 2.2. In the subsequent sections, we formally define the syntax and semantics of this evolution language. We note that the evolution

$$\mathfrak{v} ::= (n, CT, \mathtt{e})$$
$$\mathfrak{e} ::= \overline{op}$$
$$op ::= \mathtt{NewClass(C, D, \overline{C}, \overline{f})} \mid \mathtt{RenameClass(C, D)} \mid \mathtt{RenameField(C, \overline{f}, \overline{g})} \mid$$
$$\mathtt{AddField(C, \overline{D}, \overline{f}=\overline{I_{\overline{D}}})} \mid \mathtt{DeleteField(C, \overline{f})} \mid \mathtt{ChangeFieldType(C, \overline{f}, \overline{D})} \mid$$
$$\mathtt{NewSupClass(C, D, \overline{f})} \mid \mathtt{MergeClass(C, D)}$$

■ **Figure 7** Syntax of the evolution language.

$$\frac{op(\mu) \text{ defined}}{[\mu, (n, CT, \mathtt{e})] \mid op \hookrightarrow [op(\mu), (n+1, E_{op}(CT), E_{op}(\mathtt{e}))] \mid \epsilon} \quad \text{(Ev-Op)}$$

$$\frac{CT \text{ OK} \qquad \mathfrak{v} = (n, CT, \mathtt{e}) \qquad \emptyset \mid \Sigma \vdash \mathtt{e} : \mathtt{C} \text{ for some C and } \Sigma \qquad [\mu, \mathfrak{v}] \mid op \hookrightarrow [\mu', \mathfrak{v}'] \mid \epsilon}{[\mu, \mathfrak{v}] \mid op, \overline{op} \hookrightarrow [\mu', \mathfrak{v}'] \mid \overline{op}}$$

$$\text{(Ev-Cat)}$$

■ **Figure 8** Program evolution.

language itself is independent of MSVDM. The way in which goal (1) is achieved is discussed in Section 5.

First, to describe the evolution of the source code, we extend the program ($CT, \mathtt{e}$) introduced in Section 3 to be versioned; we define a *versioned program* as a triple ($n, CT, \mathtt{e}$) of a natural number $n$ representing its version, a class table $CT$, and an expression $\mathtt{e}$ corresponding to the body of the main method. Intuitively, the version number $n$ is updated when the structure of the persistent objects changes (e.g., new fields are added to the class).

The syntax of evolution language is shown in Figure 7. Program evolution is formalized as a sequence $\overline{op}$ of operations that require changes in persistent structures. Each operation $op$ corresponds to the operations triggering the database schema evolution identified above.

We formally define the evolution of the programs written in the target core language. The semantics of program evolution are given by the relation of the form $[\mu, \mathfrak{v}] \mid \mathfrak{e} \hookrightarrow [\mu', \mathfrak{v}'] \mid \mathfrak{e}'$, read as "an evolution $\mathfrak{e}$ on a versioned program $\mathfrak{v}$ under a store $\mu$ reduces to $\mathfrak{e}'$ on $\mathfrak{v}'$ under $\mu'$ (Figure 8)." In the rule Ev-Op, $op(\mu)$ denotes a store obtained by applying the evolution operation $op$ on $\mu$. Similarly, $E_{op}(CT)$ and $E_{op}(\mathtt{e})$ denote a class table that is obtained by applying the evolution operation $op$ on $CT$ and an expression obtained by $op$ on $\mathtt{e}$, respectively. As discussed later, the definition $op(\mu)$ is abstract here and will be completed in Section 5. The premise of Ev-Op indicates that such a concrete definition is necessary in our evolution language. The rule Ev-Cat describes sequential applications of $op$s. For simplicity, we do not consider any modifications that do not require changes in the persistent structures (e.g., some method's formal parameters are renamed). We do not identify the final $\mu$ and $\mathfrak{v}$ because we cannot

$$\frac{\begin{array}{c}\mathsf{CL} = \mathsf{class}\ \mathsf{C}\ \mathsf{extends}\ \mathsf{D}\ \{\ \overline{\mathsf{C}}\ \overline{\mathsf{f}};\ \}\\ \mathsf{C} \notin dom(CT) \qquad \mathsf{D} \in dom(CT)\end{array}}{E_{\mathsf{NewClass}(\mathsf{C},\mathsf{D},\overline{\mathsf{C}},\overline{\mathsf{f}})}(CT) = CT \cup \{\mathsf{C} \mapsto \mathsf{CL}\}}\ \text{(E-NewClass)}$$

$$\frac{CT(\mathsf{C}) = \mathsf{CL}}{E_{\mathsf{DeleteClass}(\mathsf{C})}(CT) = CT \setminus \{\mathsf{C} \mapsto \mathsf{CL}\}}\ \text{(E-DelClass)}$$

$$\frac{\mathsf{D} \notin dom(CT)}{E_{\mathsf{RenameClass}(\mathsf{C},\mathsf{D})}(CT) = Rename_{\mathsf{C},\mathsf{D}}(CT)}\ \text{(E-RenameClass)}$$

$$\frac{fields(\mathsf{C}) = \overline{\mathsf{C}}\ \overline{\mathsf{h}} \qquad \overline{\mathsf{h}} \cap \overline{\mathsf{g}} = \emptyset}{E_{\mathsf{RenameField}(\mathsf{C},\overline{\mathsf{f}},\overline{\mathsf{g}})}(CT) = Rename_{\mathsf{C},\overline{\mathsf{f}},\overline{\mathsf{g}}}(CT)}\ \text{(E-RenameField)}$$

$$\frac{\begin{array}{c}CT(\mathsf{C}) = \mathsf{CL} \qquad \mathsf{CL} = \mathsf{class}\ \mathsf{C}\ \mathsf{extends}\ \mathsf{D}\ \{\ \overline{\mathsf{C}}\ \overline{\mathsf{f}};\ \cdots\ \} \qquad \overline{\mathsf{g}} \cap \overline{\mathsf{f}} = \emptyset\\ \emptyset \mid \Sigma \vdash \overline{\mathsf{l}_{\overline{\mathsf{D}}}} : \overline{\mathsf{D}} \qquad \mathsf{CL}' = \mathsf{class}\ \mathsf{C}\ \mathsf{extends}\ \mathsf{D}\ \{\ \overline{\mathsf{C}}\ \overline{\mathsf{f}};\ \overline{\mathsf{D}}\ \overline{\mathsf{g}};\ \cdots\ \}\end{array}}{E_{\mathsf{AddField}(\mathsf{C},\overline{\mathsf{D}},\overline{\mathsf{g}}=\overline{\mathsf{l}_{\overline{\mathsf{D}}}})}(CT) = Expand_{\mathsf{C},\overline{\mathsf{l}_{\overline{\mathsf{D}}}}}((CL \setminus \{\mathsf{C} \mapsto \mathsf{CL}\}) \cup \{\mathsf{C} \mapsto \mathsf{CL}'\})}\ \text{(E-AddField)}$$

$$\frac{\begin{array}{c}CT(\mathsf{C}) = \mathsf{CL} \qquad \mathsf{CL} = \mathsf{class}\ \mathsf{C}\ \mathsf{extends}\ \mathsf{D}\ \{\ \overline{\mathsf{C}}\ \overline{\mathsf{f}};\ \overline{\mathsf{D}}\ \overline{\mathsf{g}};\ \cdots\ \}\\ \mathsf{CL}' = \mathsf{class}\ \mathsf{C}\ \mathsf{extends}\ \mathsf{D}\ \{\ \overline{\mathsf{D}}\ \overline{\mathsf{g}};\ \cdots\ \}\end{array}}{E_{\mathsf{DeleteField}(\mathsf{C},\overline{\mathsf{f}})}(CT) = (CT \setminus \{\mathsf{C} \mapsto \mathsf{CL}\}) \cup \{\mathsf{C} \mapsto \mathsf{CL}'\}}\ \text{(E-DelField)}$$

$$\frac{\begin{array}{c}CT(\mathsf{C}) = \mathsf{CL} \qquad \mathsf{CL} = \mathsf{class}\ \mathsf{C}\ \mathsf{extends}\ \mathsf{D}\ \{\ \overline{\mathsf{C}}\ \overline{\mathsf{f}};\ \cdots\ \}\\ \mathsf{CL}' = \mathsf{class}\ \mathsf{C}\ \mathsf{extends}\ \mathsf{D}\ \{\ \overline{\mathsf{D}}\ \overline{\mathsf{f}};\ \cdots\ \}\end{array}}{E_{\mathsf{ChangeFieldType}(\mathsf{C},\overline{\mathsf{f}},\overline{\mathsf{D}})}(CT) = (CT \setminus \{\mathsf{C} \mapsto \mathsf{CL}\}) \cup \{\mathsf{C} \mapsto \mathsf{CL}'\}}\ \text{(E-ChngFldType)}$$

$$\frac{CT(\mathsf{C}) = \mathsf{class}\ \mathsf{C}\ \mathsf{extends}\ \mathsf{E}\ \{\ \overline{\mathsf{C}}\ \overline{\mathsf{f}};\ \overline{\mathsf{D}}\ \overline{\mathsf{g}};\ \cdots\ \} \qquad CT(\mathsf{D})\ \text{undefined}}{E_{\mathsf{NewSupClass}(\mathsf{C},\mathsf{D},\overline{\mathsf{g}})}(CT) = E_{\mathsf{NewClass}(\mathsf{C},\mathsf{D},\overline{\mathsf{C}},\overline{\mathsf{f}})}(E_{\mathsf{DeleteClass}(\mathsf{C})}(E_{\mathsf{NewClass}(\mathsf{D},\mathsf{E},\overline{\mathsf{D}},\overline{\mathsf{g}})}(E_{\mathsf{DeleteField}(\mathsf{C},\overline{\mathsf{g}})}(CT))))}\ \text{(E-NewSupClass)}$$

$$\frac{\begin{array}{c}CT(\mathsf{C}) = \mathsf{class}\ \mathsf{C}\ \mathsf{extends}\ \mathsf{D}\ \{\ \overline{\mathsf{C}}\ \overline{\mathsf{f}};\ \overline{\mathsf{M}}\ \} \qquad CT(\mathsf{D}) = \mathsf{class}\ \mathsf{D}\ \mathsf{extends}\ \mathsf{E}\ \{\ \overline{\mathsf{D}}\ \overline{\mathsf{g}};\ \overline{\mathsf{M}}'\ \}\\ \mathsf{CL} = \mathsf{class}\ \mathsf{C}\ \mathsf{extends}\ \mathsf{E}\ \{\ \overline{\mathsf{D}}\ \overline{\mathsf{g}};\ \overline{\mathsf{C}}\ \overline{\mathsf{f}};\ \overline{\mathsf{M}}'\ \overline{\mathsf{M}}\ \} \qquad name(\overline{\mathsf{M}}) \cap name(\overline{\mathsf{M}}') = \emptyset\\ \mathsf{new}\ \mathsf{D}(\cdots)\ \text{does not appear in}\ CT \qquad \forall \mathsf{C}' \in dom(CT), \mathsf{C}' \neq \mathsf{C}.\mathsf{C}'\ \text{does not extends}\ \mathsf{D}\end{array}}{E_{\mathsf{MergeClass}(\mathsf{C},\mathsf{D})}(CT) = \{\mathsf{C} \mapsto \mathsf{CL}\} \cup E_{\mathsf{DeleteClass}(\mathsf{C})}(E_{\mathsf{DeleteClass}(\mathsf{D})}(CT))}\ \text{(E-MergeClass)}$$

■ **Figure 9** Class table evolution.

formally define the conditions that terminate the evolution (the initial ones are the first version of the program).

$E_{op}(CT)$ is defined in Figure 9. To simplify the definitions, we use $\mathsf{l}_\mathsf{D}$ as a shorthand for $\mathsf{new}\ \mathsf{D}(\mathsf{l}, \cdots)$ (an identifier $\mathsf{l}$ annotated with class $\mathsf{D}$ can be distinguished by the context). We also introduce an evolution operation that is not directly triggered by the developer but called only internally by the other operations (this is used to define $\mathsf{NewSupClass}$ in combination of $E_{op}$):

$$op \quad ::= \quad \cdots \mid \mathsf{DeleteClass}(\mathsf{C})$$

$$E_{op}(e) = \begin{cases} Rename_{C,D}(e) & (op = \text{RenameClass}(C,D)) \\ Rename_{C,\bar{f},\bar{g} \text{ where } \emptyset}(e) & (op = \text{RenameField}(C,\bar{f},\bar{g})) \\ Expand_{C,\bar{l}_{\overline{D}} \text{ where } \emptyset}(e) & (op = \text{AddField}(C,\overline{D},\bar{g}=\bar{l}_{\overline{D}})) \\ e & (\text{otherwise}) \end{cases}$$

■ **Figure 10** Main expression evolution

Basically, each *op* updates *CT* by adding the class declaration `CL` or fields $\bar{c}$ $\bar{f}$ to *CT* or class declaration, respectively. For simplicity, `NewClass(C,D,C̄,f̄)` creates a new class declaration with an empty method declaration list. The addition of fields also modifies the corresponding constructor invocations and those of `set` by providing the default objects $\bar{l}_{\overline{D}}$. These objects must be well-typed.[3] This modification is performed by applying the *Expand* function to the class table (Appendix F). `NewSupClass(C,D,ḡ)` extracts a new superclass from class `C` with the fields $\bar{g}$ that are originally declared in `C`. For simplicity, this superclass does not contain any method declarations. This operation is defined as a combination of `NewClass` and `DeleteField`. Furthermore, this operation replaces the superclass `E` of `C` with `D` by a combination of `DeleteClass` and `NewClass`. `MergeClass(C,D)` merges the class declaration `D` into its direct subclass `C`, making the declaration of `D` disappear from the program. This operation checks that `D` no longer appears in the program; i.e., all instance creations of `D` are removed from the source code (e.g., by replacing them with creations of `C`[4]) before applying this operation, and `C` is only the class that extends `D`.

Some evolution operations require that the entire *CT* be rewritten. For example, `RenameClass(C,D)` renames all the appearances of the class name `C` in *CT* to `D`. Similarly, `RenameField(C,f̄,ḡ)` renames all the appearances of the field names $\bar{f}$ in *CT* to $\bar{g}$. This renaming requires modifications on not only their declarations but also the code in which these class and fields are accessed. The *Rename*$_{C,D}$ function renames all the appearance of the class name `C` in *CT* to `D`. It replaces class `C` … in *CT* with class `D` … . Furthermore, it recursively applies the renaming for the bodies of all the class declarations in *CT*, and replaces all the appearance of `new C(...)` with `new D(...)`. Similarly, the *Rename*$_{C,\bar{f},\bar{g}}$ function renames all the appearances of the field names $\bar{f}$ defined in the class `C` in *CT* to $\bar{g}$. It replaces class `C` { $\bar{c}$ $\bar{f}$; … } in *CT* with class `C` { $\bar{c}$ $\bar{g}$; … }. As in the case of *Rename*$_{C,D}$, it also recursively applies the renaming for the bodies of all the class declarations in *CT*, and replaces all the appearance of `e.f`$_i$ (where `e` is given type `C`) with `e.g`$_i$. The *Expand*$_{C,\bar{l}_{\overline{D}}}$ function modifies all constructor invocations `C` and `set` invocations whose receiver type is `C` to add $\bar{l}_{\overline{D}}$ to their arguments. Their definitions are provided in Appendix F.

---

[3] To be precise, this type-checking must be performed in the *evolved* store environment (will be discussed shortly). However, in this case using $\Sigma$ for type-checking is allowed as this operation (`AddField`) does not cause any effects on $\Sigma$.

[4] This is not performed by our evolution language.

$E_{op}(\texttt{e})$ is defined in Figure 10. If $op$ is $\texttt{RenameClass(C,D)}$, all the appearances of the class name $\texttt{C}$ in $\texttt{e}$ are renamed to $\texttt{D}$ using the $Rename_{\texttt{C,D}}$ function. If $op$ is $\texttt{RenameField(C,\bar{f},\bar{g})}$, all the appearances of the field names $\bar{\texttt{f}}$ in $\texttt{e}$ are renamed to $\bar{\texttt{g}}$ using the $Rename_{\texttt{C,\bar{f},\bar{g}}}$ function. If $op$ is $\texttt{AddField(C,\bar{D},\bar{g}=\bar{l}_{\bar{D}})}$, all constructor invocation $\texttt{C}$ and $\texttt{set}$ invocations whose receiver type is $\texttt{C}$ in $\texttt{e}$ are modified using the $Expand_{\texttt{C,\bar{l}_{\bar{D}}}}$ to add $\bar{\texttt{l}}_{\bar{\texttt{D}}}$ to their arguments. Otherwise, it has no effect. We note that this rewriting of $\texttt{e}$ is not performed when $\texttt{DeleteField}$ is applied. Actually, this operation is unsafe (also discussed in Section C) and we expect a compiler to detect type errors after the evolution.

In Figure 8, the notation $op(\mu)$ denotes a store that is obtained by applying the evolution operation $op$ on $\mu$. More precisely, this is a store obtained by modifying the schema of the database $\mathscr{R}$ (denoted as $op(\mathscr{R})$), and updating the annotations for all $\texttt{l}_{\texttt{C}} \in dom(\mu)$ (because $\texttt{RenameClass(C,D)}$ may change the annotation). This is defined as follows:

$$op(\mu) = \bigcup_{\texttt{l}_{\texttt{C}} \mapsto \mathscr{R}(\texttt{C}) \in \mu} op(\texttt{l}_{\texttt{C}} \mapsto \mathscr{R}(\texttt{C}))$$

where $op(\texttt{l}_{\texttt{C}} \mapsto \mathscr{R}(\texttt{C}))$ is defined as follows:

$$op(\texttt{l}_{\texttt{C}} \mapsto \mathscr{R}(\texttt{C})) = \left\{ \begin{array}{ll} \{\texttt{l}_{\texttt{D}} \mapsto op(\mathscr{R})(\texttt{D})\} & (op = \texttt{RenameClass(C,D)}) \\ \{\texttt{l}_{\texttt{C}} \mapsto op(\mathscr{R})(\texttt{C}), \texttt{l}_{\texttt{D}} \mapsto op(\mathscr{R})(\texttt{D})\} & (op = \texttt{NewSupClass(C,D,\bar{f})}) \\ \emptyset & (op = \texttt{MergeClass(C}_0\texttt{,C)}) \\ \{\texttt{l}_{\texttt{C}} \mapsto op(\mathscr{R})(\texttt{C})\} & (\text{otherwise}) \end{array} \right.$$

We note that $\texttt{NewSupClass(C,D,\bar{f})}$ duplicates the entry, and $\texttt{MergeClass(C}_0\texttt{,C)}$ removes the entry if the annotation of the identifier is class $\texttt{C}$ whose definitions are inlined to its direct subclass $\texttt{C}_0$ in the new version.

The database with new schema $op(\mathscr{R})$ is defined for each operation, but its definition varies according to the underlying persistent object mechanism. Thus, in this section, these definitions are left abstract.

Finally, we consider the evolution of store environment $E_{op}(\Sigma)$, which is necessary to state the properties below regarding the consistency with the evolved program $(E_{op}(CT), E_{op}(\texttt{e}))$ and store $op(\mu)$. This is simply defined as:

$$E_{op}(\Sigma) = E_{op}(\bar{\texttt{l}}_{\bar{\texttt{C}}} : \bar{\texttt{C}}) = \bar{\texttt{l}}_{E_{op}(\bar{\texttt{C}})} : E_{op}(\bar{\texttt{C}})$$

where $E_{op}(\texttt{C})$ is defined as follows:

$$E_{op}(\texttt{C}) = \left\{ \begin{array}{ll} Rename_{\texttt{C}_0\texttt{,D}}(\texttt{C}) & (op = \texttt{RenameClass(C}_0\texttt{,D)}) \\ \texttt{C} & (\text{otherwise}) \end{array} \right.$$

This evolution language holds properties such as the behavior preservation. Important theorems are presented in Appendix C.

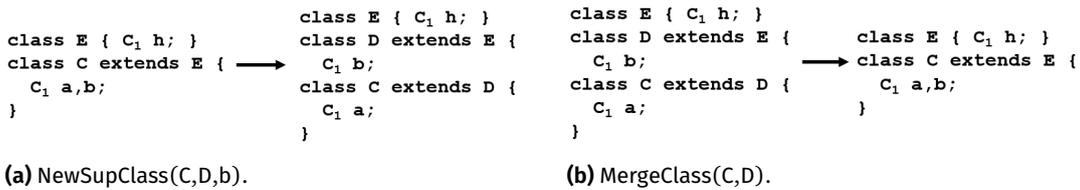

**(a)** `NewSupClass(C,D,b)`. **(b)** `MergeClass(C,D)`.

■ **Figure 11** Example of schema evolution in our SMOs: The sources and sinks of the bold arrows represent the classes before and after evolution, respectively.

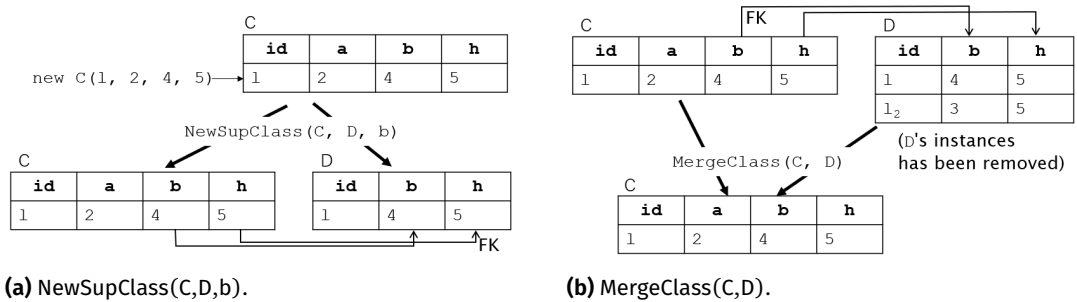

**(a)** `NewSupClass(C,D,b)`. **(b)** `MergeClass(C,D)`.

■ **Figure 12** Schema evolution shown in Figure 11 using JPA-like mapping. The sources and sinks of bold arrows denote the relations before and after evolution, respectively.

## 5 Concrete Mapping Mechanisms

The definitions of the target core language and evolution language are abstract, as the behavior of the program and required changes in the persistent structures during the evolution vary according to the specific persistent object systems. Here, we discuss two mapping mechanisms for such persistent object systems. The first one is based on a standard object-relation mapping mechanism such as JPA, in which a persistent object corresponds to a tuple in a relational database table. The second one is *signal classes* [17], where a persistent object corresponds to a timeseries relational database table, and all its update histories are stored in the table. Both mappings define the schema changes of the database, which are performed in sync with the evolution of the program. The changes in the database schema migrate the old version database to the new one. This migration is fully handled by BiDel.

We argue that these mapping mechanisms ensures that an update of persistent values occurs in a version of the program is bidirectionally propagated to other versions. This is because, for each mapping, all operations in our evolution language are directly defined using SMOs provided by BiDel, which generates the delta code to perform the bidirectional propagation. Thus, while several concrete definitions can be considered, these mappings have the advantage of instantly supporting MSVDM. We confirm that these mappings are sound by providing their formal query semantics and soundness proofs, which are provided in Appendices D and E, respectively.

$$\text{RenameClass}(C, D)(\mathcal{R}) = \text{RENAME TABLE C INTO D}, \mathcal{R}$$

$$\text{RenameField}(C, \bar{f}, \bar{g})(\mathcal{R}) = \text{RENAME COLUMN } \bar{f} \text{ IN C TO } \bar{g}, \mathcal{R}$$

$$\text{AddField}(C, \bar{D}, \bar{f} = \bar{i}_{\bar{D}})(\mathcal{R}) = \text{ADD COLUMN } \bar{f} \text{ AS } \bar{i}_{\bar{D}} \text{ INTO C}, \mathcal{R}$$

$$\text{DeleteField}(C, \bar{f})(\mathcal{R}) = \text{DROP COLUMN } \bar{f} \text{ FROM C}, \mathcal{R}$$

$$\frac{fields(D) = \bar{D} \ \bar{g}}{\text{NewClass}(C, D, \bar{C}, \bar{f})(\mathcal{R}) = \text{CREATE TABLE } C(\bar{g}, \bar{f}), \mathcal{R}}$$

$$\frac{\mathcal{R}' = \text{DROP COLUMN } \bar{f} \text{ FROM C DEFAULT } f(\bar{f}), \mathcal{R} \qquad f(\bar{f}) = \pi_{\bar{f}}(C)}{\text{ChangeFieldType}(C, \bar{f}, \bar{D}) = \text{AddField}(C, \bar{D}, \bar{f} = f(\bar{f}))(\mathcal{R}')}$$

$$\frac{\bar{f} = \mathcal{R}.C \cap \mathcal{R}.D \qquad \bar{f} \text{ are foreign keys from C to D}}{\text{MergeClass}(C, D)(\mathcal{R}) = \text{JOIN TABLE C,D INTO C}, \mathcal{R}}$$

$$\frac{CT(C) = \text{class C extends E \{ ... \}} \qquad fields(E) = \bar{E} \ \bar{h}}{\text{NewSupClass}(C, D, \bar{f})(\mathcal{R}) = \text{DECOMPOSE TABLE C INTO} C(\mathcal{R}.C), D(id, \bar{h}, \bar{f}) \text{ FK } \bar{h}, \bar{f}, \mathcal{R}}$$

■ **Figure 13**  Semantics of evolution on JPA-like mapping.

## 5.1 JPA-Like Mapping

Based on the mapping shown in Figure 4, we complete the formal definitions of the evolution operations, which are separately provided using BiDel's SMOs. Figure 12 depicts our definitions of `NewSupClass` and `MergeClass`. Assuming that the class `C` has its fields a and b, and C's superclass provides the field h,[5] `NewSupClass(C, D, b)` decomposes the corresponding table (`C`) into two tables (`C` and `D`), in which the relational schema of `C` remains the same as the original but foreign keys are set to the columns in `D` (Fig. 12a). `MergeClass(C, D)` does the inverse operation, which assumes that `C` and `D` shares the same columns and foreign keys from columns in `C` to `D` are set. As the identifier $l_2$ is no longer used in the new version, this entry does not exist in the result of `MergeClass(C, D)`.

Figure 13 defines the semantics of evolution on JPA-like mapping. The semantics of creating a new class is directly defined by the creation of a new table. The renaming of classes/fields and addition of fields are also directly defined by corresponding BiDel's SMOs. For `AddField`, we assume that the default objects $\bar{i}_{\bar{D}}$ are provided to the new

---

[5] This strategy corresponds to the JPA's `@MappedSuperclass` or the `TABLE_PER_CLASS` strategy of `@Inheritance`.

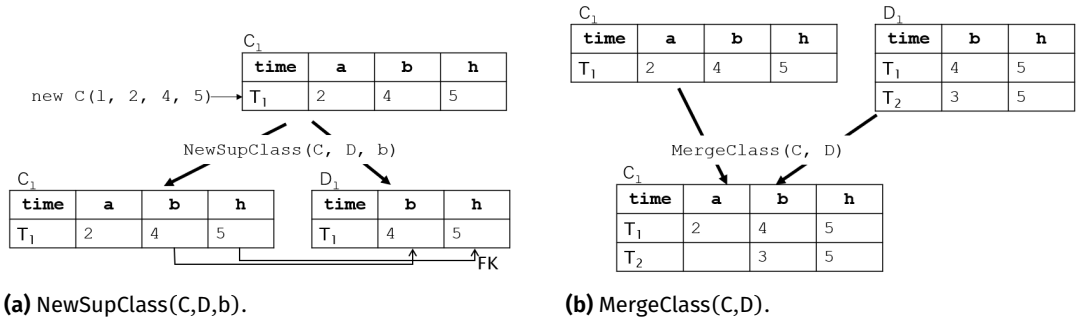

**(a)** `NewSupClass(C,D,b)`.   **(b)** `MergeClass(C,D)`.

■ **Figure 14**   Schema evolution shown in Figure 11 using signal classes. Note that each table is annotated with the label that identifies the corresponding signal class instance.

fields $\bar{f}$. `ChangeFieldType` is defined as the combination of `DeleteField` and `AddField`, where deleted column values are applied to `AddField`. `MergeClass` is defined using `JOIN TABLE`. We write the columns in `C` in $\mathscr{R}$ as $\mathscr{R}.C$. We note that `MergeClass` requires a sanity condition: foreign keys to `D`'s columns are set on `C`. These columns are used as the join keys, making the merged table consistent with the tables before merging. Similarly, for `NewSupClass`, columns corresponding to fields in common superclasses are specified as the foreign keys for the decomposition of the corresponding table.

### 5.2   Signal Classes

Based on the mapping shown in Figure 5, we complete the formal definitions of the evolution operations. Figure 14 depicts our definitions of `NewSupClass` and `MergeClass` in signal classes. While the behavior of `NewSupClass` is similar to JPA-like mapping, `MergeClass` appears to be different: `MergeClass(C,D)` performs the outer join for all pairs of $C_l$ and $D_l$ (i.e., pairs of `C` and `D` labeled with the same identifier). This is because signal classes handle time-series data where all timestamps should be preserved after join. In this outer join, values of column `a` are not provided in some rows. This missing does not raise a problem as rows from `C` must provide a value of `a` and an access to the *field* `a` results in the *latest* value of that column.

Figure 15 defines the semantics of evolution on signal classes. The introduction of a new class `C` does not cause any changes in the database schema, as database tables are always created at runtime in signal classes. The renaming of classes/fields and addition/deletion of fields are similar to the JPA-like mapping, but they require that *all* tables created from the same class to be updated using BıDEL's SMOs ($\forall C_l \in \mathscr{R}.\mathscr{R} \leftarrow$ SMO, $\mathscr{R}$ denotes a set of relations obtained by applying SMO to all tables created by the class `C`). `NewSupClass` decomposes the specified class in a similar manner by keeping the identifier l consistent. Unlike the JPA-like mapping, `MergeClass` performs the outer-join for each corresponding pair of tables identified by l.

$NewClass(C, D, \overline{C}, \overline{f})(\mathscr{R}) = \mathscr{R}$

$RenameClass(C, D)(\mathscr{R}) = \forall C_l \in \mathscr{R}.\mathscr{R} \leftarrow (\text{RENAME TABLE } C_l \text{ INTO } D_l, \mathscr{R})$

$RenameField(C, \overline{f}, \overline{g})(\mathscr{R}) = \forall C_l \in \mathscr{R}.\mathscr{R} \leftarrow (\text{RENAME COLUMN } \overline{f} \text{ IN } C_l \text{ TO } \overline{g}, \mathscr{R})$

$AddField(C, \overline{D}, \overline{f} = \overline{i_{\overline{D}}})(\mathscr{R}) = \forall C_l \in \mathscr{R}.\mathscr{R} \leftarrow (\text{ADD COLUMN } \overline{f} \text{ AS } \overline{i_{\overline{D}}} \text{ INTO } C_l, \mathscr{R})$

$DeleteField(C, \overline{f})(\mathscr{R}) = \forall C_l \in \mathscr{R}.\mathscr{R} \leftarrow (\text{DROP COLUMN } \overline{f} \text{ FROM } C_l, \mathscr{R})$

$MergeClass(C, D)(\mathscr{R}) = \forall C_l \in \mathscr{R}.\mathscr{R} \leftarrow (\text{OUTER JOIN TABLE } C_l, D_l \text{ INTO } C_l, \mathscr{R})$

$$\frac{\mathscr{R}' = \forall C_l \in \mathscr{R}.\mathscr{R} \leftarrow (\text{DROP COLUMN } \overline{f} \text{ FROM } C_l \text{ DEFAULT } f(\overline{f}), \mathscr{R}) \qquad f(\overline{f}) = \pi_{\overline{f}}(C)}{ChangeFieldType(C, \overline{f}, \overline{D}) = AddField(C, \overline{D}, \overline{f} = f(\overline{f}))(\mathscr{R}')}$$

$$\frac{CT(C) = \text{class C extends E \{ ... \}} \qquad \textit{fields}(E) = \overline{E} \ \overline{h}}{\begin{array}{l} NewSupClass(C, D, \overline{f})(\mathscr{R}) = \\ \forall C_l \in \mathscr{R}.\mathscr{R} \leftarrow \text{DECOMPOSE TABLE } C_l \text{ INTO } C_l(\mathscr{R}.C_l), D_l(\overline{h}, \overline{f}) \text{ FK } \overline{h}, \overline{f}, \mathscr{R} \end{array}}$$

■ **Figure 15**  Semantics of evolution on signal classes.

## 6  Empirical Study

To evaluate the effectiveness of our proposal, we conducted an empirical study. The aim of this study is to address the following research questions:

**RQ1.** Does our evolution language framework accurately reflect the evolution of persistent objects in existing real-world software?

**RQ2.** What software evolution operations, if any, cannot be represented in our SMOs?

For this empirical study, we utilized the evolution histories of the following open-source projects, from the first commit of any of the JPA @Entity classes to the last commit before Mar. 27th, 2024:

- Broadleaf Commerce: An e-commerce framework developed entirely in Java and built on the Spring framework.[6] The first commit occurred on Aug. 12th, 2009.
- Keycloak: An identity and access management system designed to be integrated into applications and secure services.[7] The first commit occurred on Aug. 2nd, 2013.
- Apollo Config: A configuration management system.[8] The first commit occurred on Mar. 18th, 2016.

These projects were chosen due to their diverse application areas and reliance on JPA. We did not examine the evolution of programs written using signal classes, as there is no sufficiently large codebase available for such analysis. We do not consider this lack of study is a significant flaw, as the study on signal classes' evolution can be approximated by the study on JPA applications' evolution, as discussed in Section 2.2.

---

[6] https://github.com/BroadleafCommerce/BroadleafCommerce (visited on 2024-03-27).
[7] https://github.com/keycloak/keycloak (visited on 2024-03-27).
[8] https://github.com/apolloconfig/apollo (visited on 2024-03-27).

As our SMOs are applied to each class, in this study, we examined the evolution scenarios of these programs on a per-class basis. Firstly, we utilized GitHub search to identify classes assigned the javax.persistence.Entity (jakarta.persistence.Entity) annotation within each project. For Broadleaf Commerce, we examined the top 100 matched classes along with two other classes that serve as superclasses of some matched classes; subsequently, we removed two irrelevant match results from the list of matched classes. Next, we categorized the delta of each class declaration in the program between each commit into several categories, such as non-schema evolution, schema evolution without structural changes (e.g., altering indexes and constraints), and structural schema changes (e.g., addition/deletion of columns/fields). This classification process was performed manually by one of the authors.

Under the assumption that each class can generally be modified independently (as many database schema changes only affect fields, which are unlikely to be moved to an unrelated class), the classification process primarily involved examining the delta of each class in each commit. However, we also considered two exceptional cases:

- If a class has a superclass, we considered the possibility that a "deleted" field was actually moved to the superclass. In our SMOs, this modification can be represented by firstly applying `MergeClass` and then applying `NewSupClass`. Thus, we classified this scenario as both "new inheritance" and "merge inheritance." We encountered only one instance of this modification in this study.

- A commit introducing a new base class can either be an extraction of a new superclass or a new component class. For the latter case, we also considered the possibility of field movement between the original and extracted classes in subsequent commits. Although we identified 3 cases of "new component class," we did not find any movements of fields between them.

The results of this empirical study are summarized in Table 1, which shows that a large number of modifications to JPA entity classes are not related to schema evolution. This trend is consistent with prior work [22]. "Non-code edits" refer to modifications unrelated to code (e.g., updating the copyright year in comments). "Computational changes" refer to modifications not affecting the schema (e.g., adding new methods without introducing persistent fields). "Non-structural schema evolution" refers to modifications of database constraints (e.g., changes in indexes and key constraints).

We also identified modifications that involve seven types of evolution operations (these modifications typically include computational changes, such as adding setter methods). These operations roughly correspond to our SMOs, except for the `NewClass` operation, as our focus is on differences before and after a commit. Additionally, we identified three types of evolution operations that are not supported by our SMOs.

*Answer to RQ1*. Table 1 demonstrates that a significant portion of structural schema evolution operations are classified into those supported by our SMOs. Notably, there are a large number of field-related operations. The high frequency of field deletions may be attributed to our method of counting simultaneous changes of both field name

■ **Table 1**  Classification of deltas of classes between each commit.

|  |  | Broadleaf | Keycloak | Apollo |
|---|---|---|---|---|
| # of investigated classes |  | 100 | 60 | 34 |
| Non-code edits |  | 1035 | 66 | 131 |
| Computational changes |  | 2067 | 386 | 178 |
| Non-structural schema evolution |  | 299 | 68 | 79 |
| structural schema evolution | fld(clm) addition | 315 | 189 | 24 |
|  | fld(clm) type change | 62 | 16 | 3 |
|  | fld(clm) deletion | 188 | 92 | 9 |
|  | rename fld(clm) | 52 | 26 | 3 |
|  | rename class(table) | 23 | 12 | 1 |
|  | new inheritance | 7 | 2 | 4 |
|  | merge inheritance | 3 | 3 | 1 |
| unsupported operations | change mapping | 2 | 3 | 1 |
|  | new comp. class | 3 | 0 | 0 |
|  | making flds collection | 0 | 2 | 0 |

and type as a single field deletion and another single field addition. Despite their relatively low occurrence, we also observed changes in inheritance relations in every project. In summary, these findings suggest that our SMOs are well-suited for handling existing software evolution.

We acknowledge that the results presented in Table 1 were derived syntactically, without considering the mappings discussed in Section 5. Upon examining the semantics, we discovered that each project adopts a different strategy for mapping inheritance relations to database tables: Broadleaf Commerce employs the `JOINED` strategy, while Keycloak and Apollo Config utilize the `SINGLE_TABLE` and `TABLE_PER_CLASS` strategies, respectively. Although we only discussed the `TABLE_PER_CLASS` strategy in Section 5.1, we may also consider other mappings by adjusting the semantics for `New-SupClass` and `MergeClass`. This finding underscores the usefulness of our fundamental idea of abstracting the mappings from the evolution language.

*Answer to RQ2*. Table 1 highlights some evolution operations that are not supported by our SMOs. Although these operations are not performed frequently, it is valuable to provide details about them:

**Change mapping:**  The mappings of inheritance relations to database tables can change over time. For example, the inheritance relation might initially be implemented using @MappedSuperclass, and later optimized by adopting other strategies such as `JOINED` and `SINGLE_TABLE`. This evolution scenario is not modeled in our evolution language framework and cannot be represented using our SMOs.

**New component class:**  Sometimes, a number of fields in a class, namely, `C`, are extracted to form a new class, namely, `D`, whose instance is assigned as a new field of `C`. Although this refactoring looks a simple combination of `NewClass`, `DeleteField`, and `AddField`, actually it requires additional changes in foreign key constraints. This operation can definitely be represented using a combination of BɪDᴇʟ's SMOs. We simply omitted this feature for simplicity.

**Making fields as a collection:** Another interesting case is converting fields as a key-value mapping, such as `Map<String,String> attributes`, where field names serve as keys and field values are values. This refactoring requires the field (column) names to appear in rows of a new table. Currently, this modification is not supported by BɪDᴇʟ's SMOs or other MSVDM mechanisms, thus we cannot define any mappings to BɪDᴇʟ for this scenario.

In summary, while the frequency of evolution operations unsupported by our SMOs is not significantly high, the observations outlined above highlight limitations of our evolution language framework and existing MSVDM mechanisms. These findings suggest a potential direction for future research.

*Summary of the empirical study.* Our design of SMOs is justified by the study addressing RQ1, which shows that our SMOs cover common operations in practice. Although the answer to RQ2 highlights some limitations of our SMOs, these are relatively minor (as they rarely occur in practice, as indicated by the study on RQ1). Therefore, we conclude that our SMOs function sufficiently well in practical scenarios.

*Threats to validity.* Our evolution language framework presented in Section 4 is based on a simple core language, while the real software projects used as our code base are developed using the full-set of Java. Several language mechanisms that are not supported by the core language, such as collection classes, are abstracted in this study. Therefore, this result should be considered as an approximation of software evolution in practice. Additionally, the classification of each delta is manually performed by one of the authors, who may not fully understand the intention of each modification. Different results could be produced if this classification were performed by others, although we believe that the differences would not be very significant. The details of our classification are presented in the attached file.

## 7 Related Work

MSVDM techniques are related to studies on database evolution [24, 29], e.g., schema evolution aware query language [27], and coexisting schema versions [26] such as Meta Model Management [1]. For example, the latter one helps handling of multiple schema versions after defining the new version, by allowing to derive mappings between the old version of schema and the new one. It also supports the forward propagation of queries (not backward). To our knowledge, SMOs like BɪDᴇʟ to automate the schema evolution were first proposed by PRISM [8] and its successor PRISM++ [7], and SMOs that consider coexisting schema versions (by propagating queries on a version to the different one) was first proposed by PRIMA [21]. BɪDᴇʟ is the first DEL with SMOs that is relationally complete and supports the bidirectional propagation. In particular, it is designed to fulfill the symmetric lens conditions [13]. All these studies are limited in that they only consider the database languages, and to our knowledge, our proposal is the first step to consider an application of MSVDM to persistent objects.

Studies on persistent objects have intensively been performed to improve their implementation [4] and architecture [30], and realizing them in a distributed setting [18, 19]. Among them, one of the studies related to ours is about the transparent management of persistent objects [2], which assumes that persistent objects and databases are designed and implemented separately. This technique, $M^2ORM^2$, allows both applying object-oriented design of persistent objects and using a database shared by several applications, while keeping persistence transparent in that developers do not have to know the implementation details of databases. Our proposal also supports this feature, as the mapping to the database (including the independently developed one) can separately be defined. In addition to this, our proposal supports MSVDM, which is not supported by $M^2ORM^2$. Even though our proposal only considers the evolution of persistent objects, the database schema can also evolve independently, assuming that the evolution is performed using a MSVDM technique like BiDel.

Evolution of persistent objects using SMOs is also supported by ESCHER [22], which defines its own SMOs. Like our SMOs, ESCHER's SMOs include fundamental refactoring operations such as addition/renaming of fields and changing of field types and removal of fields although, more sophisticated refactoring such as an extraction of subclass is not considered. The study on ESCHER performed an empirical study, which justifies the design of ESCHER, and we consider that this justification is also applicable to our SMOs. However, while ESCHER's empirical study focused on studying refactoring not limited to persistent objects, our study specifically focuses on persistent objects, especially JPA @Entitys. Additionally, ESCHER's SMOs do not consider MSVDM. Furthermore, ESCHER only consider the schema transformations; the runtime semantics of programming language is not considered. This means that the query semantics is fixed in ESCHER, and we cannot define the mapping from persistent objects to the database independently.

To some extent, our proposal is related to studies on dynamic software updating (DSU) [12] and similar approaches [3, 25] in that our proposal keeps the software with the old version available after releasing the new version. DSU often requires migrating data stored for running instances to the updated class definitions. This migration is not currently supported by our SMOs, as in our proposal, data migration is performed on disk rather than in main memory. However, we believe our approach can be extended to support DSU by applying a technique similar to those proposed in the literature [33, 34], which handles this migration atomically. This avenue of research is reserved for future work.

Duggan also proposed a type-correct hot swapping mechanism for running modules [9]. This mechanism allows values of different versions to be used interchangeably, enabling these values to coexist. These mechanisms ensure their type-safety; however, in some cases they allow situations where data representations inconsistent. Stoyle et al. proposed an alternative approach that forces different part of programs not to expect different representations of the same type, to keep data representations consistent [31]. Implementation issues (e.g., limited support of dynamic software updating in virtual machines) are also studied [23]. Although dynamic software updating is similar to our proposal, it is significantly different in that it is targeting to updating one instance of the running software; on the contrary, our proposal aims to allow multiple instances with different versions to coexist while keeping data shared by them consistent.

MSVDM, which allows multiple schema versions to coexist in a database system, reminds us a software system where different versions of packages coexist. In general, multiple versions of a package are incompatible in that they provide different sets of definitions (e.g., function names and data structures), or different behaviors under the same definition. $\lambda_{VL}$ is a calculus that is designed as a basis for the idea of *programming with versions* [32], which allows for incompatible versions to coexist and if they are accessed, the appropriate version is selected from the contexts. This approach is different from that of MSVDM. In $\lambda_{VL}$, a value is modeled as a versioned value that have multiple raw values tagged with versions, while MSVDM shares the same database status among different versions.

## 8   Conclusion

This paper proposes the evolution language framework that realizes MSVDM for persistent objects, where concrete definitions of persistent objects can be provided independently. This framework includes the abstract core language for persistent objects and the abstract evolution language that manages the evolution of them. These languages support multiple persistent object mechanisms that are independently developed, which is demonstrated by defining two different persistent object mechanisms, namely, JPA-like mapping and signal classes. In these definitions, every evolution operation is defined by BiDel, ensuring that the database status is shared between the old version and the new one, which is the important property of MSVDM.

This is the first step to realize MSVDM for persistent objects, revoking several future research directions. First, we can enrich the core language to enable representation of more sophisticated queries. One fundamental property regarding the expressiveness of queries is its relational completeness. We can enjoy this property if we can directly express queries using relational algebra. However, this requires a knowledge about the details of the database schema, which is abstract in our setting. Thus, it is necessary to design an abstract query language where mappings from persistent objects to the database are defined independently. Additionally, our empirical study revealed potential future research directions on enhancing both our evolution language framework and existing MSVDM mechanisms. Finally, designing tool supports based on our proposal is another possible research topic.

**Acknowledgements**   This study was supported by KAKENHI 21H03418 and 24K02922.

## A   Remaining Definitions of Abstract Core Language

In this section, we complete the definitions of the core language introduced in Section 3. First, we assume that $CT(\texttt{C}) = \texttt{class C} \dots$ for any $\texttt{C} \in dom(CT)$, $\texttt{Object} \notin dom(CT)$, and there are no cycles in the inheritance relations. We also assume that all fields and

$$\frac{Select(\mu, \mathsf{l}_\mathsf{C}, \mathsf{f}) = \mathsf{l}' \qquad \mathsf{l}'_{\mathsf{C}'} \in dom(\mu)}{\mu \mid \mathsf{l}_\mathsf{C}.\mathsf{f} \longrightarrow \mu \mid \mathsf{l}'_{\mathsf{C}'}} \text{(R-Field)}$$

$$\frac{mbody(\mathsf{m}, \mathsf{C}) = \overline{\mathsf{x}}.\mathsf{e} \qquad \mathsf{m} \neq \mathsf{set}}{\mu \mid \mathsf{l}_\mathsf{C}.\mathsf{m}(\overline{\mathsf{l}_{\overline{\mathsf{C}}}}) \longrightarrow \mu \mid \mathsf{e}[\overline{\mathsf{x}}/\overline{\mathsf{l}_{\overline{\mathsf{C}}}}, \mathsf{this}/\mathsf{l}_\mathsf{C}]} \text{(R-Invk)}$$

$$\mu \mid \mathsf{new}\ \mathsf{C}(\overline{\mathsf{l}, \mathsf{l}_{\overline{\mathsf{C}}}}) \longrightarrow Insert(\mu, \mathsf{C}, \mathsf{l}, \overline{\mathsf{l}}) \mid \mathsf{l}_\mathsf{C} \text{(R-New)}$$

$$\mu \mid \mathsf{l}_\mathsf{C}.\mathsf{set}(\overline{\mathsf{l}_{\overline{\mathsf{C}}}}) \longrightarrow Update(\mu, \mathsf{l}_\mathsf{C}, \overline{\mathsf{l}}) \mid \mathsf{l}_\mathsf{C} \text{(R-Set)}$$

■ **Figure 16**  Small-step computation rules for expressions.

methods in the same class, and all parameters in the same method are distinct. In the following discussion, we use the following auxiliary definitions: $fields(\mathsf{C})$, $mbody(\mathsf{m}, \mathsf{C})$, and $mtype(m, C)$. Based on the declaration class $\mathsf{C}$ extends $\mathsf{D}\ \{\overline{\mathsf{C}}\ \overline{\mathsf{f}}; \overline{\mathsf{M}}\}$, those definitions are provided as follows:

- $fields(\mathsf{C}) = \begin{cases} \cdot & (\mathsf{C} = \mathsf{Object}) \\ fields(\mathsf{D}), \overline{\mathsf{C}}\ \overline{\mathsf{f}} & (otherwise) \end{cases}$

- $mbody(\mathsf{m}, \mathsf{C}) = \begin{cases} \overline{\mathsf{x}}.\mathsf{e}_0 & (\mathsf{C}_0\ \mathsf{m}(\overline{\mathsf{C}}\ \overline{\mathsf{x}})\ \{\ \mathsf{return}\ \mathsf{e}_0;\ \} \in \overline{\mathsf{M}}) \\ mbody(\mathsf{m}, \mathsf{D}) & (otherwise) \end{cases}$

- $mtype(\mathsf{m}, \mathsf{C}) = \begin{cases} \overline{\mathsf{C}} \rightarrow \mathsf{C}_0 & (\mathsf{C}_0\ \mathsf{m}(\overline{\mathsf{C}}\ \overline{\mathsf{x}})\ \{\ \mathsf{return}\ \mathsf{e}_0;\ \} \in \overline{\mathsf{M}}) \\ mtype(\mathsf{m}, \mathsf{D}) & (otherwise) \end{cases}$

### A.1  Runtime Semantics

We show the reduction rules of expressions in Figure 16. These are given by the relation of the form $\mu \mid \mathsf{e} \longrightarrow \mu' \mid \mathsf{e}'$, which is read as "an expression $\mathsf{e}$ under a store $\mu$ reduces to $\mathsf{e}'$ under $\mu'$." The store $\mu$ is a set of mapping $\mathsf{l}_\mathsf{C} \mapsto \mathscr{R}(\mathsf{C})$, where $\mathsf{l}_\mathsf{C}$ is the identifier of the persistent object, and $\mathscr{R}(\mathsf{C})$ is a relation that contains the persistent object identified by $\mathsf{l}_\mathsf{C}$.

Each reduction rule is defined in terms of interactions with the database: $Select$ (selection from the database), $Insert$ (insertion of a persistent object into the database), and $Update$ (update of the persistent object in the database). $Select$ returns a value $\mathsf{l}'_{\mathsf{C}'}$ from the store $\mu$ using the field $\mathsf{f}$ and its receiver $\mathsf{l}_\mathsf{C}$. $Insert$ and $Update$ update the store $\mu$. Thus, a field access results in a query on $\mu$ using the field, and an instance creation results in an insertion of a database entity into $\mu$. We note that the class tags assigned to identifiers are removed when they are inserted into the database. This annotation is recovered when the identifier $\mathsf{l}$ is selected from the database (the most specific $\mathsf{C}$ in $\forall \mathsf{l}_\mathsf{C} \in dom(\mu)$ is selected). Here, all these definitions are abstract; concrete definitions are given when we discuss the mapping mechanisms of specific persistent objects in Section 5.

There are also congruence rules that enable reductions of subexpressions, which are omitted in this paper.

$$\frac{x : C \in \Gamma}{\Gamma \mid \emptyset \vdash x : C} \quad \text{(T-Var)} \qquad \frac{\Sigma(l_C) = C}{\emptyset \mid \Sigma \vdash l_C : C} \quad \text{(T-Id)} \qquad \frac{\Sigma(l_C) = C}{\emptyset \mid \Sigma \vdash l : C} \quad \text{(T-Id1)}$$

$$\frac{\Gamma \mid \Sigma \vdash e_0 : C_0 \qquad \mathit{fields}(C_0) = \overline{C}\ \overline{f}}{\Gamma \mid \Sigma \vdash e_0.f_i : C_i} \quad \text{(T-Field)}$$

$$\frac{\mathit{fields}(C) = \overline{C}\ \overline{f} \qquad \Gamma \mid \Sigma \vdash \overline{e} : \overline{D} \qquad \overline{D} <: \overline{C} \qquad \Gamma \mid \Sigma \vdash l : C}{\Gamma \mid \Sigma \vdash \text{new } C(l, \overline{e}) : C} \quad \text{(T-New)}$$

$$\frac{\Gamma \mid \Sigma \vdash e_0 : C_0 \qquad \mathit{fields}(C_0) = \overline{C}\ \overline{f} \qquad \Gamma \mid \Sigma \vdash \overline{e} : \overline{D} \qquad \overline{D} <: \overline{C}}{\Gamma \mid \Sigma \vdash e_0.\text{set}(\overline{e}) : C_0} \quad \text{(T-Set)}$$

$$\frac{\Gamma \mid \Sigma \vdash e_0 : C_0 \qquad \mathit{mtype}(m, C_0) = \overline{C} \to C \qquad \Gamma \mid \Sigma \vdash \overline{e} : \overline{D} \qquad \overline{D} <: \overline{C}}{\Gamma \mid \Sigma \vdash e_0.m(\overline{e}) : C} \quad \text{(T-Invk)}$$

■ **Figure 17** Expression typing.

$$\frac{\overline{x} : \overline{C}, \text{this} : C \mid \emptyset \vdash e_0 : D_0 \qquad D_0 <: C_0}{C_0\ m(\overline{C}\ \overline{x})\ \{\ \text{return } e_0;\ \}\ \text{OK IN } C} \quad \text{(T-Method)}$$

$$\frac{\overline{M} \text{ OK IN } C \qquad \forall m \in \overline{M}. \mathit{mtype}(m, D) \text{ is undefined.}}{\text{class } C \text{ extends } D\ \{\ \overline{C}\ \overline{f};\ \overline{M}\ \}\ \text{OK}} \quad \text{(T-Class)}$$

■ **Figure 18** Method and class typing.

### A.2 Type System

This section presents the type system of the target core language. First, we present the subtyping relation as a reflective and transitive closure of the immediate subclass relation given by the extends clause of the class declaration:

$$C <: C \qquad \frac{C <: D \qquad D <: E}{C <: E} \qquad \frac{\text{class } C \text{ extends } D\{ \dots \}}{C <: D}$$

Typing rules for expressions are shown in Figure 17. Typing rules for *CT* (i.e., class and method declarations) are shown in Figure 18. All these typing rules are straightforward adaptation of the typing rules in FJ.

To formulate the type soundness, we must also define the relationship between the static store environment $\Sigma$ with the runtime database $\mu$. We say that $\mu$ is wellformed with respect to a store environment $\Sigma$ if $\mu$ is consistent with $\Sigma$; i.e., every relation (table) in $\mu$ associated with a class $C$ has an attribute for every field of $C$ and stores an object identifier of the field's type. This relation $\Sigma \vdash \mu$ is formally defined as follows:

$$\frac{\forall l_C \mapsto \mathscr{R}(C) \in \mu.fields(C) = \overline{C}\ \overline{f} \wedge \emptyset \mid \Sigma \vdash \pi_{f_i}(\mathscr{R}(C)) : D_i \qquad \overline{D} <: \overline{C}}{\Sigma \vdash \mu}$$

<div align="right">(T-Store)</div>

We also assume that a program $(CT, e)$ does not contain identifiers annotated with the class name, i.e., only raw identifiers (in the constructor invocations) appear in the program.

## B  Properties of the Abstract Core Language

The type soundness of the target core language can be described by the following theorems.

**Theorem B.1** (Preservation). *If* $\Gamma \mid \Sigma \vdash e : C$, $\Sigma \vdash \mu$, *and* $\mu \mid e \longrightarrow \mu' \mid e'$, *then* $\Gamma \mid \Sigma \vdash e' : C'$ *and* $\Sigma \vdash \mu'$ *for some* $C' <: C$.

**Theorem B.2** (Progress). *Let* $\Gamma \mid \Sigma \vdash e : C$ *for some* $\Gamma$ *and* $\Sigma$. *Then, either* $e$ *is an identifier* $l_{C'}$ *or, for any* $\mu$ *such that* $\Sigma \vdash \mu$, *there are some* $e'$ *and* $\mu'$ *such that* $\mu \mid e \longrightarrow \mu' \mid e'$.

To prove these theorems, it is necessary to provide the concrete definitions of *Select*, *Insert*, and *Update* in Figure 16. In other words, we can derive the properties that these definitions should ensure, and we formulate them as the following lemmas. In Section 5, we provide the concrete definitions of them in which these lemmas can be proved. Assuming that all those lemmas hold, the proofs of the theorems are outlined in Appendix E.

**Lemma B.1.** *If* $\Gamma \mid \Sigma \vdash l_{C_0}.f : C$, $\Gamma \mid \Sigma \vdash l_{C_0} : C_0$, *fields*$(C_0) = \overline{C}\ \overline{f}$ *where* $C_i = C$ *and* $f_i = f$, *and* $\Sigma \vdash \mu$, *then there is some* $l'_C$ *such that* $Select(\mu, l_{C_0}, f_i) = l'$, $l'_C \in dom(\mu)$, $\forall l'_{C'} \in dom(\mu).C' <: C$, *and* $\Gamma \mid \Sigma \vdash l'_C : C$.

**Lemma B.2.** *If* $\Gamma \mid \Sigma \vdash \overline{l}_{\overline{C}} : \overline{C}$, $\Gamma \mid \Sigma : C \vdash new\ C(l, \overline{l}_{\overline{C}}) : C$, *and* $\Sigma \vdash \mu$, *then* $\Sigma \vdash Insert(\mu, C, l, \overline{l})$.

**Lemma B.3.** *If* $\Gamma \mid \Sigma \vdash \overline{l}_{\overline{C}} : \overline{C}$, $\Gamma \mid \Sigma \vdash l_C.set(\overline{l}_{\overline{C}}) : C$, *and* $\Sigma \vdash \mu$, *then* $\Sigma \vdash Update(\mu, l_C, \overline{l})$.

## C  Properties of the Evolution Language

One desirable property that our evolution language should ensure is that the proposed SMOs preserve type safety. However, the `DeleteField` operation appears to be unsafe because there may be accesses to the deleted field in the program, and our evolution language does not handle fixing such errors. Similarly, the `ChangeFieldType` operation is unsafe. We consider that error detection is the compiler's responsibility; developers should be responsible for addressing them.

In other words, we can show that other SMO's are type safe; i.e., if an evolution operation is applied to a wellformed program, then the resulting program is also wellformed. This fact is formally described as the following theorem (its proof is outlined in Appendix E):

**Theorem C.1.** *If CT OK, $\Sigma \vdash \mu$, and $\Gamma \mid \Sigma \vdash e : C$ for some C, then $E_{op}(CT)$ OK (op is neither DeleteClass, ChangeFieldType, and DeleteField), $E_{op}(\Sigma) \vdash op(\mu)$, and $\Gamma \mid E_{op}(\Sigma) \vdash E_{op}(e) : E_{op}(C)$.*

Next, we provide the theorem regarding the behavior preservation, i.e., our evolution operations preserve the behavior of the program.

**Theorem C.2.** *If $\Gamma \mid \Sigma \vdash e : C$, $\Sigma \vdash \mu$, $[\mu, (n, CT, e)] \mid op \hookrightarrow [op(\mu), (n + 1, E_{op}(CT), E_{op}(e))] \mid \epsilon$, and $\mu \mid e \longrightarrow \mu' \mid e'$, then if $\Gamma \mid E_{op}(\Sigma) \vdash E_{op}(e) : E_{op}(C)$, then $E_{op}(\Sigma) \vdash op(\mu)$ and $op(\mu) \mid E_{op}(e) \longrightarrow op(\mu') \mid E_{op}(e')$.*

Again, to prove these theorems, it is necessary to provide the concrete definitions of $op(\mathcal{R})$, but we can derive the properties that those definitions should provide. We formulate them as the following lemmas. The concrete definitions provided in Section 5 prove these lemmas. Assuming that all these lemmas hold, the proof of the theorem is outlined in Appendix E.

**Lemma C.1.** *Suppose $Select(\mu, l_C, f) = l'$, $\Gamma \mid \Sigma \vdash l_C : C$, and $fields(C) = \bar{C} \, \bar{f}$ for some $\Gamma$. Then, (1) If $op = \text{RenameField}(C, \bar{f}, \bar{g})$ and $f = f_i$, then $Select(op(\mu), l_C, g_i) = l'$; (2) otherwise, $Select(op(\mu), l_C, f) = l'$.*

**Lemma C.2.** *(1) $\text{RenameClass}(C, D)(Insert(\mu, C, l, \bar{l})) = Insert(\text{RenameClass}(C, D)(\mu), D, l, \bar{l})$; (2) if $op \neq \text{RenameClass}$ or $op = \text{RenameClass}(C_0, D)$ and $C \neq C_0$, then $op(Insert(\mu, C, l, \bar{l})) = Insert(\, op(\mu), C, l, \bar{l})$.*

**Lemma C.3.** *$op(Update(\mu, l_C, \bar{l})) = Update(op(\mu), l_C, \bar{l})$*

## D  Concrete Query Semantics

We complete the formal definitions of the persistent objects' runtime semantics with the two different mapping mechanisms, JPA-like mapping and signal classes.

### D.1  JPA-Like Mapping

Figure 19 shows an example of the semantics of interactions to the database during the computation defined in Figure 16. We assume that each relation in $\mathcal{R}$ has the attribute id, which is the primary key, and the identifier l given in the constructor invocation is assigned to this attribute. In this setting, $\mu$ is a set of mappings from an identifier $l_C$ to the relation whose schema name is C in which a tuple whose id value is l is inserted. As introduced in Section 2.1, $\oplus$ denotes a destructive update in which the right-hand side of the mapping is updated. Thus, $Insert(\mu, C, l, \bar{l})$ results in a store in which the relation C is updated by inserting the tuple $(l, \bar{l})$. If the instance of

$$Select(\mu, l_C, f) = \pi_f(\sigma_{id=l}(\mu(l_C))) \qquad \frac{\sigma_{id=l}(\mu(l_C)) \neq \emptyset}{Insert(\mu, C, l, \bar{l}) = \mu}$$

$$\frac{\sigma_{id=l}(\mu(l_C)) = \emptyset}{Insert(\mu, C, l, \bar{l}) = \mu \oplus \{l_C \mapsto (\mu(l) \cup (l, \bar{l}))\}}$$

$$Update(\mu, l_C, \bar{l}) = \mu \oplus \{l_C \mapsto (\mu(l_C) \oplus (l, \bar{l}))\}$$

■ **Figure 19** Semantics of query on JPA-like mapping.

$$Select(\mu, l_C, f) = \pi_f(\sigma_{latest}(\mu(l_C))) \qquad \frac{\mu(l_C) \neq \emptyset}{Insert(\mu, C, l, \bar{l}) = \mu}$$

$$\frac{\mu(l_C) = \emptyset}{Insert(\mu, C, l, \bar{l}) = \mu \cup \{l_C \mapsto C_l(\bot, \bar{l})\}}$$

$$\frac{t > \sigma_{latest}(\pi_{time}(\mu(l_C)))}{Update(\mu, l_C, \bar{l}) = \mu \oplus \{l_C \mapsto (\mu(l_C) \cup (t, \bar{l}))\}}$$

■ **Figure 20** Semantics of query on signal classes.

$l_C$ already exists, it has no effect. Similarly, $Update(\mu, l_C, \bar{l})$ results in a store in which the relation C is updated by replacing the tuple identified by $l$ with $(l, \bar{l})$ (we override $\oplus$ to denote a destructive update of a relation in which the tuple identified by the primary key is updated).

In these settings, we can prove the lemmas introduced in Appendix C. The proofs are shown in Appendix E.

### D.2 Signal Classes

Query semantics of signal classes reflect the lifecycle model proposed by Kamina et al. [17]. An instance creation creates a corresponding time-series table if it does not exist, and an access to a field results in the latest value of the time-series table.[9]

Figure 20 shows how the interactions to the database behave in signal classes. We assume that each relation $C \in \mathscr{R}$ has the attribute time that contains the timestamp indicating the time at which the tuple is inserted. $Select(\mu, l_C, f)$ results in the value of the attribute in the *latest* tuple in $\mu(l_C)$. We use the predicate *latest*, which is true only if the *time* field of the tuple has the largest value among the relations.

---

[9] Signal classes also provide an operation called snapshot to set the time cursor of the specified instance to the specified timestamp. In this proposal, this feature is abstracted, as it does not interact with the schema evolution.

Unlike the JPA-like mapping, the identifier $l$ is used to identify the table $C_l$, which is created when the constructor $C$ is invoked with the identifier $l$ (and this is the first time that the constructor is invoked). $Update(\mu, l_C, \bar{l})$ inserts a new tuple $(t, \bar{l})$ with a fresh timestamp $t$ into $\mu(l_C)$. We note that these semantics do not perform the UPDATE statement of SQL, as all the values updating the current value are treated as an insertion into the time-series table with a new timestamp.

In these settings, we can prove the lemmas introduced in Appendix C. The proofs are shown in Appendix E.

## E Proofs

### E.1 Proofs of Theorems

*Proof of Theorem B.1.* By induction on the derivation of $\mu \mid e \longrightarrow \mu' \mid e'$.

Case R-FIELD:

$$e = l_{C_0}.f \qquad e' = l'_C \qquad Select(\mu, l_{C_0}, f) = l'$$

By the assumption, $\Gamma \mid \Sigma \vdash l_{C_0}.f : C_0$. By T-FIELD, we obtain $\Gamma \mid \Sigma \vdash l_{C_0} : C_0$ and $fields(C_0) = \bar{C}\ \bar{f}$ where $C_i = C$ and $f_i = f$. By Lemma B.1, we obtain $\Gamma \mid \Sigma \vdash Select(\mu, l_{C_0}, f) : C$, i.e., $\Gamma \mid \Sigma \vdash l' : C$, completing the case.

Case R-NEW:

$$e = \text{new } C(l, \bar{l_{\bar{C}}}) \qquad \mu' = Insert(\mu, C, l, \bar{l}) \qquad e' = l_C$$

By the assumption, $\Gamma \mid \Sigma \vdash \text{new } C(l, \bar{l}) : C$. By T-NEW, we obtain $\Gamma \mid \Sigma \vdash l_C : C$. By Lemma B.2, we obtain $\Sigma \vdash Insert(\mu, C, l, \bar{l})$, completing the case.

Case R-SET:

$$e = l_C.\text{set}(\bar{l_{\bar{C}}}) \qquad \mu' = Update(\mu, l_C, \bar{l}) \qquad e' = l_C$$

By the assumption, $\Gamma \mid \Sigma \vdash l_C.\text{set}(\bar{l_{\bar{C}}}) : C$. By T-SET, we obtain $\Gamma \mid \Sigma \vdash l_C : C$. By Lemma B.3, we obtain $\Sigma \vdash Update(\mu, l_C, \bar{l})$, completing the case.

Case R-INVK:

$$e = l_{C_0}.m(\bar{l_{\bar{C}}}) \qquad mbody(m, C_0) = \bar{x}.e_0$$
$$e' = e_0[\bar{x}/\bar{l_{\bar{C}}}, \text{this}/l_{C_0}]$$

By the assumption, $\Gamma \mid \Sigma \vdash l_{C_0}.m(\bar{l_{\bar{C}}}) : C$. By T-INVK, we obtain $\Gamma \mid \Sigma \vdash l_{C_0} : C_0$ and $mtype(m, C_0) = \bar{C} \rightarrow C$. By the definitions of $mbody$ and $mtype$, we can easily show that there is $D <: C$ and $\Gamma, \bar{x} : \bar{C}, \text{this} : C_0 \mid \Sigma \vdash e_0 : D$. We can also easily show that the substitution lemma holds in the calculus; thus we obtain $\Gamma \mid \Sigma \vdash e_0[\bar{x}/\bar{l_{\bar{C}}}, \text{this}/l_{C_0}] : E$ and $E <: D$, completing the case.

$\square$

*Proof of Theorem B.2.* By induction on the derivation of $\Gamma \mid \Sigma \vdash e : C$.

Case T-Var: Cannot occur.

Cases T-Id and T-Idi: Immediate.

Case T-Field: $e = e_0.f$.

If $e_0$ is not an identifier, the induction hypothesis and R-Cong complete the case. Otherwise, let $e_0 = l_C$. By Lemma B.1, there is some $l'$ such that $Select(\mu, l_C, f) = l'$. Then, R-Field completes the case.

Case T-New: $e = \text{new } C(l, \overline{e})$

If some of $\overline{e}$ are not an identifier, the induction hypothesis and R-Cong complete the case. Otherwise, R-New completes the case.

Case T-Set: $e = e_0.\text{set}(\overline{e})$

If $e_0$ is not an identifier, the induction hypothesis and R-Cong complete the case. Otherwise, if some of $\overline{e}$ are not an identifier, the induction hypothesis and R-Cong complete the case. Otherwise, let $e = l_C.\text{set}(\overline{l_{\overline{C}}})$. Then, R-Set completes the case.

Case T-Invk: $e = e_0.\text{m}(\overline{e})$

If $e_0$ is not an identifier, the induction hypothesis and R-Cong complete the case. Otherwise, if some of $\overline{e}$ are not an identifier, the induction hypothesis and R-Cong complete the case. Otherwise, let $e = l_C.\text{m}(\overline{l_{\overline{C}}})$. Then, R-Invk completes the case. $\square$

*Proof of Theorem C.1.* We must prove the theorem for each case of $op$.

Case $\text{NewClass}(C_0, D, \overline{C}, \overline{f})$: By E-NewClass, we obtain $E_{op}(CT)$ OK and $\Gamma \mid op(\Sigma) \vdash e : C$ immediately.

Case $\text{NewSupClass}(C_0, D, \overline{g})$: By induction on $\Gamma \mid \Sigma \vdash e : C$, we easily obtain $\Gamma \mid op(\Sigma) \vdash e : C$ (we note that results of $mtype(m, C)$ and $fields(C)$ in the derivation using $CT$ remain the same in the derivation using $op(CT)$). Similarly, for each method declaration $C' \text{ m}(\overline{C} \ \overline{x})\{ \text{ return } e'; \}$ in $C''$ contained in $CT$, we obtain $\overline{x} : \overline{C}, \text{this} : C'' \mid op(\Sigma) \vdash e' : C'$. Thus, we obtain $E_{op}(CT)$ OK.

Case $\text{MergeClass}(C_0, D)$: Similar to the case of $\text{NewSupClass}(C_0, D, \overline{g})$.

Case $\text{RenameClass}(C_0, D)$: By induction on $\Gamma \mid \Sigma \vdash e : C$ and the definition of $Rename_{C_0, D}$, we obtain $\Gamma \mid op(\Sigma) \vdash Rename_{C_0, D}(e) : op(C)$. Similarly, for each method declaration $C' \text{ m}(\overline{C} \ \overline{x})\{ \text{ return } e'; \}$ in $C''$ contained in $CT$, we obtain $\overline{x} : \overline{C}, \text{this} : C'' \mid op(\Sigma) \vdash Rename_{C_0, D}(e') : op(C')$. Thus, we obtain $E_{op}(CT)$ OK.

Case $\text{RenameField}(C_0, \overline{f}, \overline{g})$: By induction on $\Gamma \vdash e : C$ and the definition of $Rename_{C_0, \overline{f}, \overline{g} \text{ where } \Gamma}$, we obtain $\Gamma \mid op(\Sigma) \vdash Rename_{C_0, \overline{f}, \overline{g} \text{ where } \emptyset}(e) : op(C)$. Similarly, for each method declaration $C' \text{ m}(\overline{C} \ \overline{x})\{ \text{ return } e'; \}$ in $C''$ contained in $CT$, we obtain:

$$\overline{x} : \overline{C}, \text{this} : C'' \mid op(\Sigma) \vdash Rename_{C_0, \overline{f}, \overline{g} \text{ where } \overline{x} : \overline{C}, \text{this} : C''}(e') : op(C').$$

Thus, we obtain $E_{op}(CT)$ OK.

Case $\text{AddField}(C_0, \overline{D}, \overline{g} = \overline{l_{\overline{C}}})$: By $\Gamma \mid \Sigma \vdash \overline{l_{\overline{D}}} : \overline{D}$, induction on $\Gamma \mid \Sigma \vdash e : C$, and the definition of $Expand_{C, \overline{l_{\overline{D}}} \text{ where } \Gamma}$, we obtain $\Gamma \mid op(\Sigma) \vdash Expand_{C_0, \overline{l} \text{ where } \emptyset}(e) : op(C)$. Similarly, for each method declaration $C' \text{ m}(\overline{C} \ \overline{x})\{ \text{ return } e'; \}$ in $C''$ contained in $CT$, we obtain:

$$\overline{x} : \overline{C}, \text{this} : C'' \mid op(\Sigma) \vdash Expand_{C_0, \overline{l_{\overline{D}}} \text{ where } \overline{x} : \overline{C}, \text{this} : C''}(e') : op(C').$$

Thus, we obtain $E_{op}(CT)$ OK.

Case $\text{ChangeFieldType}(C_0, \overline{f}, \overline{D})$: By induction on $\Gamma \mid \Sigma \vdash e : C$ and E-ChngFldType, we obtain $\Gamma \mid op(\Sigma) \vdash E_{op(e)} : E_{op}(C)$. Similarly, we obtain $E_{op}(CT)$ OK. $\square$

*Proof of Theorem C.2.* It is easy to show that $\Gamma \mid E_{op}(\Sigma) \vdash E_{op}(e) : E_{op}(C)$ and $E_{op}(\Sigma) \vdash op(\mu)$ for each $op$. Below, we show $op(\mu) \mid E_{op}(e) \longrightarrow op(\mu') \mid E_{op}(e')$ by induction on the derivation of $\mu \mid e \longrightarrow \mu' \mid e'$.

Case R-FIELD:

$$e = l_c.f \qquad e' = l'_{c'} \qquad Select(\mu, l_c, f) = l'$$

If $op$ is not RenameField, $E_{op}(l_c.f) = l_c.f$. By Lemma C.1, we obtain $Select(op(\mu), l_c, f) = l'$. We easily obtain that there is some $C'$ such that $l'_{c'} \in op(\mu)$ and $\forall l'_{c''} \in dom(op(\mu)).C' <: C''$. Thus, by R-FIELD, $op(\mu) \mid l_c.f \longrightarrow op(\mu') \mid l'_{c'}$, completing this subcase. If $op$ is RenameField, $E_{op}(l_c.f) = Rename_{C,\bar{f},\bar{g} \text{ where } \emptyset}(l_c.f)$. By T-FIELD, $\Gamma \mid \Sigma \vdash l_c : C$ and $fields(C) = \overline{C} \, \overline{f}$. If $f = f_i$, $E_{op}(l_c.f) = l_c.g_i$. By Lemma C.1, we obtain $Select(RenameField(C, \bar{f}, \bar{g})(\mu), l_c, g_i) = l'$. Thus, R-FIELD completes this subcase. Otherwise, $E_{op}(l_c.f) = l_c.f$. Thus, Lemma C.1 and R-FIELD complete this case.

Case R-NEW:

$$e = \text{new } C(l, \bar{l}_{\bar{c}}) \qquad \mu' = Insert(\mu, C, l, \bar{l}) \qquad e' = l_c$$

If $op$ is not RenameClass, $E_{op}(\text{new } C(l, \bar{l}_{\bar{c}})) = \text{new } C(l, \bar{l}_{\bar{c}})$. By Lemma C.2, $op(Insert(\mu, C, l, \bar{l})) = Insert(op(\mu), C, l, \bar{l})$. Thus, R-NEW completes this subcase. If $op$ is RenameClass($C_0, D$), there are two subcases. If $C_0 \neq C$, $E_{RenameClass(C_0,D)}(\text{new } C(l, \bar{l}_{\bar{c}})) = \text{new } C(l, \bar{l}_{\bar{c}})$. Thus, Lemma C.2 and R-NEW complete the case. If $C_0 = C$, $E_{RenameClass(C_0,D)}(\text{new } C(l, \bar{l}_{\bar{c}})) = \text{new } D(l, \bar{l}_{\bar{c}})$. By Lemma C.2, RenameClass($C, D$)($Insert(\mu, C, l, \bar{l})$) = $Insert$(Rename- Class($C, D$)($\mu$), $D, l, \bar{l}$). Thus, R-NEW completes the case.

Case R-SET:

$$e = l_c.\text{set}(\bar{l}_{\bar{c}}) \qquad \mu' = Update(\mu, l_c, \bar{l}) \qquad e' = l_c$$

$E_{op}(l_c.\text{set}(\bar{l}_{\bar{c}})) = l_c.\text{set}(\bar{l}_{\bar{c}})$. By Lemma C.3, $op(Update(\mu, l_c, \bar{l})) = Update(op(\mu), l_c, \bar{l})$. Thus, R-SET completes the case.

Case R-INVK:

$$e = l_c.m(\bar{l}_{\bar{c}}) \qquad mbody(m, C) = \bar{x}.e_0$$
$$e' = e_0[\bar{x}/\bar{l}_{\bar{c}}, \text{this}/l_c]$$

$E_{op}(l_c.m(\bar{l}_{\bar{c}})) = l_c.m(\bar{l}_{\bar{c}})$. For each $op$, by the definition of $mbody$, we can check that $mbody(m, op(C_0)) = \bar{x}.e$ on the evolved class table $E_{op}(CT)$. Thus, R-INVK completes the case.

$\square$

## E.2  Proofs of Lemmas in the Context of Java Persistent API

*Proof of Lemma B.1.* By the assumption $\Gamma \mid \Sigma \vdash l_{c_0}.f : C$, we obtain that $\mu(l_{c_0}) \neq \emptyset$, i.e., $\mu(l_{c_0}) = D_0$ for some $D_0$. By the assumption of the lemma, T-STORE, and the uniqueness of $l_{c_0}$, we obtain $D_0 = C_0$. Thus, by T-STORE, $\pi_f(\sigma_{id=l_0}(C_0))$ must not be $\emptyset$. Let $\pi_f(\sigma_{id=l_0}(C_0)) = l'$. By induction on the length of the reduction, we also obtain that there is some $C'$ such that $l'_{C'} \in dom(\mu)$ (intuitively, $l'$ in the store must be used as an argument for constructor invocation during the preceding reductions). By the fact that the duplicated entries in $\mu$ are created only by NewSupClass, there are no cycles in the inheritance relations, and T-Store, we obtain $\Gamma \mid \Sigma \vdash l'_{C'} : C$, completing the proof. □

*Proof of Lemma B.2.* By T-New, we obtain $\mathit{fields}(C) = \overline{D} \, \overline{f}$, and $\overline{C} <: \overline{D}$. If $\sigma_{id=l}(\mu(l_C)) = \emptyset$, by T-Store we obtain $\Sigma \vdash \mathit{Insert}(\mu, C, l, \overline{l})$. The other case is immediate. □

*Proof of Lemma B.3.* Similar to the proof of Lemma B.2. □

*Proof of Lemma C.1.* $\mathit{Select}(\mu, l_C, f) = \pi_f(\sigma_{id=l}(\mu(l_C))) = l'$.

Case (1): Let the relation schema $\mu(l_C) = C(\cdots, f, \cdots)$. As $\mathit{fields}(C) = \overline{C} \, \overline{f}$ and $f = f_i$, we obtain RenameField$(C, \overline{f}, \overline{g})(\mu)(l_C) = C(\cdots, g_i, \cdots)$. Thus, $\pi_{g_i}(\sigma_{id=l}(\mathrm{RenameField}(C, \overline{f}, \overline{g})(\mu)(l_C))) = l'$, completing the case.

Case (2): We need to check for all the cases of $op$. We only show the case of NewSupClass$(C, D, \overline{f})$. First, we obtain $\mu(l_C) \neq \emptyset$, and thus NewSupClass$(C, D, \overline{f})(\mu)(l_C) = C$ for some $C_0$. Thus, we obtain NewSupClass$(C, D, \overline{f})(\mu)(l_C) = \mu(l_C)$, completing the case. □

*Proof of Lemma C.2.* Case (1) can be proved by the following equations (we write the relation obtained by renaming the name $C$ of $\mu(l_C)$ to $D$ as $\mathit{Rename}_{C,D}(\mu(l_C))$):

$$\begin{aligned}
\text{RenameClass}&(C, D)(\mathit{Insert}(\mu, C, l, \overline{l})) \\
&= \text{RenameClass}(C, D)(\mu \oplus \{l_C \mapsto (\mu(l_C) \cup (l, \overline{l}))\}) \\
&= \mu \oplus \{l_C \mapsto (\mathit{Rename}_{C,D}(\mu(l_C)) \cup (l, \overline{l}))\} \\
&= \mu \oplus \{l_C \mapsto (\mathit{Rename}_{C,D}(\mu(l_C)) \cup (l, \overline{l}))\} \\
&= \mathit{Insert}(\text{RenameClass}(C, D)(\mu), D, l, \overline{l})
\end{aligned}$$

Case (2): Similar to the proof of case (1). □

*Proof of Lemma C.3.* Similar to the proof of case (2) of Lemma C.2. □

### E.3 Proofs of Lemmas in the Context of Signal Classes

*Proof of Lemma B.1.* By induction on the length of the reduction, we obtain $\mu(l_{C_0}) = D_l$ for some $D$. By the assumption of the lemma, T-Store, and the uniqueness of $l_{C_0}$, we obtain $D = C_0$. Thus, by T-Store, we easily show that $\Gamma \mid \Sigma \vdash \pi_f(\sigma_{latest}(\mu(l_{C_0}))) : C_0$ and $\pi_f(\sigma_{latest}(\mu(l_{C_0})))$ must not be $\emptyset$. Let $\pi_f(\sigma_{latest}(\mu(l_{C_0}))) = l'$. By induction on the length of the reduction and the fact that the duplicated entries in $\mu$ are created only by NewSupClass, we also obtain that there is some $C'$ such that $l'_{C'} \in dom(\mu)$ and $C' <: C$, completing the proof. □

*Proof of Lemma B.2.* The case in which $\mu(l_C) \neq \emptyset$ is immediate. If $\mu(l_C) = \emptyset$, by T-New we obtain $\mathit{fields}(C) = \overline{D} \, \overline{f}$, and $\overline{C} <: \overline{D}$. Thus, by T-Store, we obtain $\Sigma \vdash \mathit{Insert}(\mu, C, l, \overline{l})$. □

*Proof of Lemma B.3.* As $\mu(\mathsf{l_c}) \neq \emptyset$, we obtain $\mu(\mathsf{l_c}) = \mathsf{C}$. Thus, by T-Set, we obtain $\mathit{fields}(\mathsf{C}) = \overline{\mathsf{D}}\ \overline{\mathsf{f}}$, and $\overline{\mathsf{C}} <: \overline{\mathsf{D}}$. Thus, by T-Store, we obtain $\Sigma \vdash \mathit{Update}(\mu, \mathsf{l_c}, \overline{\mathsf{l}})$. $\qquad\square$

*Proof of Lemma C.1.* $\mathit{Select}(\mu, \mathsf{l_c}, \mathsf{f}) = \pi_{\mathsf{f}}(\sigma_{\mathit{latest}}(\mu(\mathsf{l_c}))) = \mathsf{l'}$.

Case (1): Let the relation schema $\mu(\mathsf{l_c}) = \mathsf{C_l}(\cdots, \mathsf{f}, \cdots)$. As $\mathit{fields}(\mathsf{C}) = \overline{\mathsf{C}}\ \overline{\mathsf{f}}$ and $\mathsf{f} = \mathsf{f_i}$, we obtain $\mathsf{RenameField}(\mathsf{C_l}, \overline{\mathsf{f}}, \overline{\mathsf{g}})(\mu)(\mathsf{l_c}) = \mathsf{C_l}(\cdots, \mathsf{g_i}, \cdots)$. Thus, $\pi_{\mathsf{g_i}}(\sigma_{\mathit{latest}}(\mathsf{RenameField}(\mathsf{C}, \overline{\mathsf{f}}, \overline{\mathsf{g}})(\mu)(\mathsf{l_c}))) = \mathsf{l'}$, completing the case.

Case (2): We need to check for all the cases of *op*. We only show the case of $\mathsf{MergeClass}(\mathsf{C}, \mathsf{D})$.

Case $\mathsf{MergeClass}(\mathsf{C}, \mathsf{D})$: Let $\mathsf{MergeClass}(\mathsf{C}, \mathsf{D}) = \mathsf{C_0}$. There are following subcases. If $\mathsf{C_0} = \mathsf{C}$ or $\mathsf{C_0} = \mathsf{D}$, by the semantics of $\mathsf{MergeClass}$ on signal classes, we have $\mu(\mathsf{l_c}) \subseteq \mathsf{MergeClass}(\mathsf{C}, \mathsf{D})\ (\mu)(\mathsf{l_c})$. Thus, we obtain $\mathit{Select}(\mathsf{MergeClass}(\mathsf{C}, \mathsf{D})(\mu), \mathsf{l_c}, \mathsf{f}) = \mathsf{l'}$, completing the case. Otherwise, obviously $\mathsf{MergeClass}(\mathsf{C}, \mathsf{D})(\mu)(\mathsf{l_c}) = \mu(\mathsf{l_c})$, completing the case. $\qquad\square$

*Proof of Lemma C.2.* If $\mu(\mathsf{l_c}) = \emptyset$, case (1) can be proved by the following equations (the case $\mu(\mathsf{l_c}) \neq \emptyset$ is trivial):

$$
\begin{aligned}
&\mathsf{RenameClass}(\mathsf{C}, \mathsf{D})(\mathit{Insert}(\mu, \mathsf{C}, \mathsf{l}, \overline{\mathsf{l}})) \\
&\quad = \mathsf{RenameClass}(\mathsf{C}, \mathsf{D})(\mu \cup \{\mathsf{l_c} \mapsto \mathsf{C_l}(\bot, \overline{\mathsf{l}})\}) \\
&\quad = \mathsf{RenameClass}(\mathsf{C}, \mathsf{D})(\mu) \cup \{\mathsf{l_c} \mapsto \mathsf{D_l}(\bot, \overline{\mathsf{l}})\} \\
&\quad = \mathit{Insert}(\mathsf{RenameClass}(\mathsf{C}, \mathsf{D})(\mu), \mathsf{D}, \mathsf{l}, \overline{\mathsf{l}})
\end{aligned}
$$

Case (2): The case $\mu(\mathsf{l_c}) \neq \emptyset$ is trivial. If $\mu(\mathsf{l_c}) = \emptyset$, we need to check for all the cases of *op*. We only show the case of $\mathsf{NewSupClass}(\mathsf{C_0}, \mathsf{D}, \overline{\mathsf{f}})$. By the semantics of $\mathsf{NewSupClass}$ on signal classes, we obtain $\mathsf{NewSupClass}(\mathsf{C_0}, \mathsf{D}, \overline{\mathsf{f}})(\mu)(\mathsf{l_c}) = \mu(\mathsf{l_c})$. Thus, the following equations prove the lemma:

$$
\begin{aligned}
&\mathsf{NewSupClass}(\mathsf{C_0}, \mathsf{D}, \overline{\mathsf{f}})(\mathit{Insert}(\mu, \mathsf{C}, \mathsf{l}, \overline{\mathsf{l}})) \\
&\quad = \mathsf{NewSupClass}(\mathsf{C_0}, \mathsf{D}, \overline{\mathsf{f}})(\mu \cup \{\mathsf{l_c} \mapsto \mathsf{C_l}(\bot, \overline{\mathsf{l}})\}) \\
&\quad = \mathsf{NewSupClass}(\mathsf{C_0}, \mathsf{D}, \overline{\mathsf{f}})(\mu) \cup \{\mathsf{l_c} \mapsto \mathsf{C_l}(\bot, \overline{\mathsf{l}})\} \\
&\quad = \mathit{Insert}(\mathsf{NewSupClass}(\mathsf{C_0}, \mathsf{D}, \overline{\mathsf{f}})(\mu), \mathsf{C}, \mathsf{l}, \overline{\mathsf{l}})
\end{aligned}
$$

$\qquad\square$

*Proof of Lemma C.3.* Similar to the proof of Lemma C.2. $\qquad\square$

## F   Auxiliary Definitions

We provide several auxiliary definitions, which are referred by the definitions in Section 4.

# Evolution Language Framework for Persistent Objects

## F.1 Definitions of Class Renaming

$$Rename_{C,D} = \bigcup_{C' \in dom(CT)} \{C' \mapsto Rename_{C,D}(CT(C'))\}$$

$Rename_{C,D}(\text{class } C' \text{ extends } D \, \{\, \overline{C} \; \overline{f}; \overline{M} \,\}) =$
$\qquad \text{class } Rename_{C,D}(C') \text{ extends } Rename_{C,D}(D) \, \{\, Rename_{C,D}(\overline{C}) \; \overline{f}; \; Rename_{C,D}(\overline{M}) \,\}$

$Rename_{C,D}(C' \; m(\overline{C} \; \overline{x})\{\text{return } e;\}) =$
$\qquad Rename_{C,D}(C') \; m(Rename_{C,D}(\overline{C}) \; \overline{x})\{\text{return } Rename_{C,D}(e);\}$

$Rename_{C,D}(x) = x$

$Rename_{C,D}(l) = l$

$Rename_{C,D}(e.f) = Rename_{C,D}(e).f$

$Rename_{C,D}(e.\text{set}(\overline{e})) = Rename_{C,D}(e).\text{set}(Rename_{C,D}(\overline{e}))$

$Rename_{C,D}(e.m(\overline{e})) = Rename_{C,D}(e).m(Rename_{C,D}(\overline{e}))$

$Rename_{C,D}(\text{new } C'(l,\overline{e})) = \text{new } Rename_{C,D}(C')(l, Rename_{C,D}(\overline{e}))$

$$Rename_{C,D}(C_0) = \begin{cases} D & (C_0 = C) \\ C_0 & (\text{otherwise}) \end{cases}$$

## F.2 Definitions of Field Renaming

$$Rename_{C,\overline{f},\overline{g}} = \bigcup_{C \in dom(CT)} \{C \mapsto Rename_{C,\overline{f},\overline{g}}(CT(C))\}$$

$$\frac{C = C_0}{Rename_{C,\overline{f},\overline{g}}(\text{class } C_0 \text{ extends } D\{\overline{C} \; \overline{f}; \overline{M}\}) = \text{class } C_0 \text{ extends } D\{\overline{C} \; \overline{g}; \; Rename_{C,\overline{f},\overline{g} \text{ in } C_0}(\overline{M})\}}$$

$$\frac{C \neq C_0}{Rename_{C,\overline{f},\overline{g}}(\text{class } C_0 \text{ extends } D\{\overline{C} \; \overline{f}; \overline{M}\}) = \text{class } C_0 \text{ extends } D\{\overline{C} \; \overline{f}; \; Rename_{C,\overline{f},\overline{g} \text{ in } C_0}(\overline{M})\}}$$

$Rename_{C,\overline{f},\overline{g} \text{ in } C_0}(C \; m(\overline{C} \; \overline{x})\{\text{return } e;\}) = C \; m(\overline{C} \; \overline{x})\{\text{return } Rename_{C,\overline{f},\overline{g} \text{ where this}:C_0, \overline{x}:\overline{C}}(e);\}$

$Rename_{C,\overline{f},\overline{g} \text{ where } \Gamma}(x) = x$

$Rename_{C,\overline{f},\overline{g} \text{ where } \Gamma}(l) = l$

$Rename_{C,\overline{f},\overline{g} \text{ where } \Gamma}(e.\text{set}(\overline{e})) = Rename_{C,\overline{f},\overline{g} \text{ where } \Gamma}(e).\text{set}(Rename_{C,\overline{f},\overline{g} \text{ where } \Gamma}(\overline{e}))$

$Rename_{C,\overline{f},\overline{g} \text{ where } \Gamma}(e.m(\overline{e})) = Rename_{C,\overline{f},\overline{g} \text{ where } \Gamma}(e).m(Rename_{C,\overline{f},\overline{g} \text{ where } \Gamma}(\overline{e}))$

$Rename_{C,\overline{f},\overline{g} \text{ where } \Gamma}(\text{new } C(l,\overline{e})) = \text{new } C(l, Rename_{C,\overline{f},\overline{g} \text{ where } \Gamma}(\overline{e}))$

$$\frac{\Gamma \mid \Sigma \vdash e : C \qquad \mathit{fields}(C) = \overline{C} \; \overline{f}}{Rename_{C,\overline{f},\overline{g} \text{ where } \Gamma}(e.f_i) = Rename_{C,\overline{f},\overline{g} \text{ where } \Gamma}(e).g_i}$$

$$\frac{\Gamma \mid \Sigma \vdash e : D \qquad D \neq C}{Rename_{C,\bar{f},\bar{g} \text{ where } \Gamma}(e.f) = Rename_{C,\bar{f},\bar{g} \text{ where } \Gamma}(e).f}$$

### F.3 Definitions for Expanding the Use Site of Constructor and set

$$Expand_{C,\bar{l}_{\bar{D}}} = \bigcup_{C \in dom(CT)}\{C \mapsto Expand_{C,\bar{l}_{\bar{D}}}(CT(C))\}$$

$$Expand_{C,\bar{l}_{\bar{D}}}(\text{class } C_0 \text{ extends } D\{\overline{C}\ \overline{f};\ \overline{M}\}) = \text{class } C_0 \text{ extends } D\{\overline{C}\ \overline{g};\ Expand_{C,\bar{l}_{\bar{D}} \text{ in } C_0}(\overline{M})\}$$

$$Expand_{C,\bar{l}_{\bar{D}} \text{ in } C_0}(C\ m(\overline{C}\ \overline{x})\{\text{return } e;\}) = C\ m(\overline{C}\ \overline{x})\{\text{return } Expand_{C,\bar{l}_{\bar{D}} \text{ where } this:C_0,\bar{x}:\overline{C}}(e);\}$$

$$Expand_{C,\bar{l}_{\bar{D}} \text{ where } \Gamma}(x) = x$$

$$Expand_{C,\bar{l}_{\bar{D}} \text{ where } \Gamma}(l) = l$$

$$Expand_{C,\bar{l}_{\bar{D}} \text{ where } \Gamma}(e.m(\overline{e})) = Expand_{C,\bar{l}_{\bar{D}} \text{ where } \Gamma}(e).m(Expand_{C,\bar{l}_{\bar{D}} \text{ where } \Gamma}(\overline{e}))$$

$$Expand_{C,\bar{l}_{\bar{D}} \text{ where } \Gamma}(\text{new } C(l,\overline{e})) = \text{new } C(l, Expand_{C,\bar{l}_{\bar{D}} \text{ where } \Gamma}(\overline{e}), \bar{l}_{\bar{D}})$$

$$Expand_{C,\bar{l}_{\bar{D}} \text{ where } \Gamma}(e.f) = Expand_{C,\bar{l}_{\bar{D}} \text{ where } \Gamma}(e).f$$

$$\frac{\Gamma \mid \Sigma \vdash e : C}{Expand_{C,\bar{l}_{\bar{D}} \text{ where } \Gamma}(e.\text{set}(\overline{e})) = Expand_{C,\bar{l}_{\bar{D}} \text{ where } \Gamma}(e).\text{set}(Expand_{C,\bar{l}_{\bar{D}} \text{ where } \Gamma}(\overline{e}), \bar{l}_{\bar{D}})}$$

$$\frac{\Gamma \mid \Sigma \vdash e : C_0 \qquad C_0 \neq C}{Expand_{C,\bar{l}_{\bar{D}} \text{ where } \Gamma}(e.\text{set}(\overline{e})) = Expand_{C,\bar{l}_{\bar{D}} \text{ where } \Gamma}(e).\text{set}(Expand_{C,\bar{l}_{\bar{D}} \text{ where } \Gamma}(\overline{e}))}$$

$$\frac{C_0 \neq C}{Expand_{C,\bar{l}_{\bar{D}} \text{ where } \Gamma}(\text{new } C_0(l,\overline{e})) = \text{new } C_0(l, Expand_{C,\bar{l}_{\bar{D}} \text{ where } \Gamma}(\overline{e}))}$$

## About the authors

**Tetsuo Kamina** is an Associate Professor at Oita University. His research interests include the design and implementation of programming languages, programming practices, and programming education. You can contact him at kamina@oita-u.ac.jp.
https://orcid.org/0000-0003-0288-1908

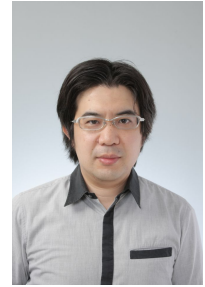

**Tomoyuki Aotani** is an Associate Professor at Sanyo Onoda City University. His research interests focus on the principles and mechanisms of programming languages, with a particular emphasis on modularity. You can contact him at aotani@rs.socu.ac.jp.
https://orcid.org/0000-0003-4538-0230

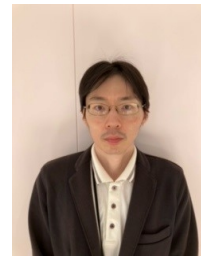

**Hidehiko Masuhara** is a Professor of Mathematical and Computing Science at Institute of Science Tokyo (formerly Tokyo Institute of Technology). His research interest is programming languages, especially on aspect- and context-oriented programming, partial evaluation, computational reflection, meta-level architectures, parallel/concurrent computing, and programming environments. Contact him at masuhara@acm.org.
https://orcid.org/0000-0002-8837-5303

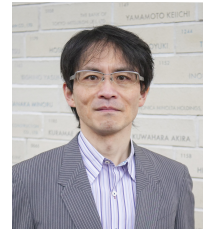